\input harvmac.tex
\input epsf.tex
\input amssym
\input ulem.sty
\input graphicx.tex
\input amssym

\def\({\left(}
\def\){\right)}
\def\[{\left[}
\def\]{\right]}

\def\<{\langle}
\def\>{\rangle}


\let\includefigures=\iftrue
\let\useblackboard=\iftrue
\newfam\black

\def\figin{\epsfcheck\figin}\def\figins{\epsfcheck\figins}
\def\epsfcheck{\ifx\epsfbox\UnDeFiNeD
\message{(NO epsf.tex, FIGURES WILL BE IGNORED)}
\gdef\figin##1{\vskip2in}\gdef\figins##1{\hskip.5in}
\else\message{(FIGURES WILL BE INCLUDED)}%
\gdef\figin##1{##1}\gdef\figins##1{##1}\fi}
\def\DefWarn#1{}
\def\figinsert{\goodbreak\midinsert}
\def\ifig#1#2#3{\DefWarn#1\xdef#1{fig.~\the\figno}
\writedef{#1\leftbracket fig.\noexpand~\the\figno} %
\figinsert\figin{\centerline{#3}}\medskip\centerline{\vbox{\baselineskip12pt
\advance\hsize by -1truein\noindent\footnotefont{\bf
Fig.~\the\figno:} #2}}
\bigskip\endinsert\global\advance\figno by1}


\includefigures
\message{If you do not have epsf.tex (to include figures),}
\message{change the option at the top of the tex file.}
\input epsf
\def\figin{\epsfcheck\figin}\def\figins{\epsfcheck\figins}
\def\epsfcheck{\ifx\epsfbox\UnDeFiNeD
\message{(NO epsf.tex, FIGURES WILL BE IGNORED)}
\gdef\figin##1{\vskip2in}\gdef\figins##1{\hskip.5in}
\else\message{(FIGURES WILL BE INCLUDED)}%
\gdef\figin##1{##1}\gdef\figins##1{##1}\fi}
\def\DefWarn#1{}
\def\figinsert{\goodbreak\midinsert}
\def\ifig#1#2#3{\DefWarn#1\xdef#1{fig.~\the\figno}
\writedef{#1\leftbracket fig.\noexpand~\the\figno}%
\figinsert\figin{\centerline{#3}}\medskip\centerline{\vbox{
\baselineskip12pt\advance\hsize by -1truein
\noindent\footnotefont{\bf Fig.~\the\figno:} #2}}
\endinsert\global\advance\figno by1}
\else
\def\ifig#1#2#3{\xdef#1{fig.~\the\figno}
\writedef{#1\leftbracket fig.\noexpand~\the\figno}%
\global\advance\figno by1} \fi

\def\figin{\epsfcheck\figin}\def\figins{\epsfcheck\figins}
\def\epsfcheck{\ifx\epsfbox\UnDeFiNeD
\message{(NO epsf.tex, FIGURES WILL BE IGNORED)}
\gdef\figin##1{\vskip2in}\gdef\figins##1{\hskip.5in}
\else\message{(FIGURES WILL BE INCLUDED)}%
\gdef\figin##1{##1}\gdef\figins##1{##1}\fi}
\def\DefWarn#1{}
\def\figinsert{\goodbreak\midinsert}
\def\ifig#1#2#3{\DefWarn#1\xdef#1{fig.~\the\figno}
\writedef{#1\leftbracket fig.\noexpand~\the\figno} %
\figinsert\figin{\centerline{#3}}\medskip\centerline{\vbox{\baselineskip12pt
\advance\hsize by -1truein\noindent\footnotefont{\bf
Fig.~\the\figno:} #2}}
\bigskip\endinsert\global\advance\figno by1}

\def \eps {\epsilon}
\def \la {\langle}
\def \ra {\rangle}
\def \p {\partial}

\def \e {\epsilon}
\def \b {\bar}
\def \t {\widetilde}

\lref\CollierEXN{
  S.~Collier, Y.~Gobeil, H.~Maxfield and E.~Perlmutter,
  ``Quantum Regge Trajectories and the Virasoro Analytic Bootstrap,''
[arXiv:1811.05710 [hep-th]].
}

\lref\DasCNV{
  D.~Das, S.~Datta and S.~Pal,
  ``Universal asymptotics of three-point coefficients from elliptic representation of Virasoro blocks,''
Phys.\ Rev.\ D {\bf 98}, no. 10, 101901 (2018).
[arXiv:1712.01842 [hep-th]].
}

\lref\KomargodskiEK{
  Z.~Komargodski and A.~Zhiboedov,
JHEP {\bf 1311}, 140 (2013).
[arXiv:1212.4103 [hep-th]].
}

\lref\FitzpatrickYX{
  A.~L.~Fitzpatrick, J.~Kaplan, D.~Poland and D.~Simmons-Duffin,
  ``The Analytic Bootstrap and AdS Superhorizon Locality,''
JHEP {\bf 1312}, 004 (2013).
[arXiv:1212.3616 [hep-th]].
}

\lref\RychkovIQZ{
  S.~Rychkov,
  ``EPFL Lectures on Conformal Field Theory in D $\geq$ 3 Dimensions,''
[arXiv:1601.05000 [hep-th]].
}

\lref\SimmonsDuffinGJK{
  D.~Simmons-Duffin,
  ``The Conformal Bootstrap,''
[arXiv:1602.07982 [hep-th]].
}

\lref\PolandEPD{
  D.~Poland, S.~Rychkov and A.~Vichi,
  ``The Conformal Bootstrap: Theory, Numerical Techniques, and Applications,''
[arXiv:1805.04405 [hep-th]].
}

\lref\AldayNJK{
  L.~F.~Alday,
  ``Large Spin Perturbation Theory for Conformal Field Theories,''
Phys.\ Rev.\ Lett.\  {\bf 119}, no. 11, 111601 (2017).
[arXiv:1611.01500 [hep-th]].
}

\lref\MazacQEV{
  D.~Mazac,
  ``Analytic bounds and emergence of AdS$_{2}$ physics from the conformal bootstrap,''
JHEP {\bf 1704}, 146 (2017).
[arXiv:1611.10060 [hep-th]].
}

\lref\MazacMDX{
  D.~Mazac and M.~F.~Paulos,
  ``The Analytic Functional Bootstrap I: 1D CFTs and 2D S-Matrices,''
[arXiv:1803.10233 [hep-th]].
}

\lref\FerraraYT{
  S.~Ferrara, A.~F.~Grillo and R.~Gatto,
  ``Tensor representations of conformal algebra and conformally covariant operator product expansion,''
Annals Phys.\  {\bf 76}, 161 (1973)..
}

\lref\PolyakovGS{
  A.~M.~Polyakov,
  ``Nonhamiltonian approach to conformal quantum field theory,''
Zh.\ Eksp.\ Teor.\ Fiz.\  {\bf 66}, 23 (1974), [Sov.\ Phys.\ JETP {\bf 39}, 9 (1974)]..
}

\lref\CollierCLS{
  S.~Collier, Y.~H.~Lin and X.~Yin,
  ``Modular Bootstrap Revisited,''
[arXiv:1608.06241 [hep-th]].
}

\lref\RattazziPE{
  R.~Rattazzi, V.~S.~Rychkov, E.~Tonni and A.~Vichi,
  ``Bounding scalar operator dimensions in 4D CFT,''
JHEP {\bf 0812}, 031 (2008).
[arXiv:0807.0004 [hep-th]].
}

\lref\AldayMF{
  L.~F.~Alday and J.~M.~Maldacena,
  ``Comments on operators with large spin,''
JHEP {\bf 0711}, 019 (2007).
[arXiv:0708.0672 [hep-th]].
}

\lref\HellermanNRA{
  S.~Hellerman, D.~Orlando, S.~Reffert and M.~Watanabe,
  ``On the CFT Operator Spectrum at Large Global Charge,''
JHEP {\bf 1512}, 071 (2015).
[arXiv:1505.01537 [hep-th]].
}

\lref\MoninJMO{
  A.~Monin, D.~Pirtskhalava, R.~Rattazzi and F.~K.~Seibold,
  ``Semiclassics, Goldstone Bosons and CFT data,''
JHEP {\bf 1706}, 011 (2017).
[arXiv:1611.02912 [hep-th]].
}

\lref\LashkariVGJ{
  N.~Lashkari, A.~Dymarsky and H.~Liu,
  ``Eigenstate Thermalization Hypothesis in Conformal Field Theory,''
[arXiv:1610.00302 [hep-th]].
}

\lref\PappadopuloJK{
  D.~Pappadopulo, S.~Rychkov, J.~Espin and R.~Rattazzi,
  ``OPE Convergence in Conformal Field Theory,''
Phys.\ Rev.\ D {\bf 86}, 105043 (2012).
[arXiv:1208.6449 [hep-th]].
}

\lref\Ingham{
  A.E.~Ingham,
  ``A Tauberian theorem for partitions,''
Ann.\ of\ Math. (2) {\bf 42}, 1075-1090 (1941).
}

\lref\Subhankulov{
  M.~A.~Subhankulov, F.~I.~An,
  ``Complex tauberian theorems for the one-sided and two-sided Stieltjes transform,''
1974 Math. USSR Izv. 8 145.
}

\lref\SubhankulovR{
  M.~A.~Subhankulov,
  ``Tauberian Theorems with Remainder'', ({\it in Russian})
1976, Izdat. Nauka, Moscow.
}

\lref\Korevaar{
  J.~Korevaar,
  ``Tauberian theory: a century of development.,''
Springer, 2004. 497 p.
}

\lref\DasCNV{
  D.~Das, S.~Datta and S.~Pal,
  ``Modular crossings, OPE coefficients and black holes,''
[arXiv:1712.01842 [hep-th]].
}

\lref\CardyIE{
  J.~L.~Cardy,
  ``Operator Content of Two-Dimensional Conformally Invariant Theories,''
Nucl.\ Phys.\ B {\bf 270}, 186 (1986)..
}

\lref\Postnikov{
A.~G.~Postnikov,
``Tauberian theory and its applications,''
  American Mathematical Soc. ,
  144, 1980
  }

\lref\QiaoXIF{
  J.~Qiao and S.~Rychkov,
  ``A tauberian theorem for the conformal bootstrap,''
JHEP {\bf 1712}, 119 (2017).
[arXiv:1709.00008 [hep-th]].
}

{ \overfullrule=0pt
\lref\MukhametzhanovZJA{
  B.~Mukhametzhanov and A.~Zhiboedov,
  ``Analytic Euclidean Bootstrap'',
\break
[arXiv:1808.03212, [hep-th]].
}
}

\lref\DiFrancescoNK{
  P.~Di Francesco, P.~Mathieu and D.~Senechal,
  ``Conformal Field Theory,''
  Springer, 1997. 890 p.
}

\lref\DasVEJ{
  D.~Das, S.~Datta and S.~Pal,
  ``Charged structure constants from modularity,''
JHEP {\bf 1711}, 183 (2017).
[arXiv:1706.04612 [hep-th]].
}

\lref\CaronHuotVEP{
  S.~Caron-Huot,
  ``Analyticity in Spin in Conformal Theories,''
JHEP {\bf 1709}, 078 (2017).
[arXiv:1703.00278 [hep-th]].
}

\lref\CaronHuotNS{
  S.~Caron-Huot,
  ``Asymptotics of thermal spectral functions,''
Phys.\ Rev.\ D {\bf 79}, 125009 (2009).
[arXiv:0903.3958 [hep-ph]].
}

\lref\RychkovLCA{
  S.~Rychkov and P.~Yvernay,
  ``Remarks on the Convergence Properties of the Conformal Block Expansion,''
Phys.\ Lett.\ B {\bf 753}, 682 (2016).
[arXiv:1510.08486 [hep-th]].
}

\lref\HeemskerkPN{
  I.~Heemskerk, J.~Penedones, J.~Polchinski and J.~Sully,
  ``Holography from Conformal Field Theory,''
JHEP {\bf 0910}, 079 (2009).
[arXiv:0907.0151 [hep-th]].
}

\lref\JafferisZNA{
  D.~Jafferis, B.~Mukhametzhanov and A.~Zhiboedov,
  ``Conformal Bootstrap At Large Charge,''
JHEP {\bf 1805}, 043 (2018).
[arXiv:1710.11161 [hep-th]].
}

\lref\StromingerEQ{
  A.~Strominger,
  ``Black hole entropy from near horizon microstates,''
JHEP {\bf 9802}, 009 (1998).
[hep-th/9712251].
}

\lref\Deutsch{
J.~M.~Deutsch, 
``Quantum statistical mechanics in a closed system," 
Physical Review A, {\bf 43(4)}, 2046 (1991). 
}

\lref\Srednicki{
M.~Srednicki,
``Chaos and quantum thermalization,"
Physical Review E, {\bf 50(2)}, 888 (1994).
[cond-mat/9403051].
}

\lref\Rigol{
M.~Rigol, V.~Dunjko and M.~Olshanii, 
``Thermalization and its mechanism for generic isolated quantum systems," 
Nature, {\bf 452(7189)}, 854-858 (2008).
[arXiv:0708.1324 [cond-mat.stat-mech]].
}

\lref\DAlessioRWT{
  L.~D'Alessio, Y.~Kafri, A.~Polkovnikov and M.~Rigol,
  ``From quantum chaos and eigenstate thermalization to statistical mechanics and thermodynamics,''
Adv.\ Phys.\  {\bf 65}, no. 3, 239 (2016).
[arXiv:1509.06411 [cond-mat.stat-mech]].
}

\lref\ZamolodchikovIE{
  Al.~B.~Zamolodchikov,
  ``Conformal Symmetry In Two-dimensions: An Explicit Recurrence Formula For The Conformal Partial Wave Amplitude,''
Commun.\ Math.\ Phys.\  {\bf 96}, 419 (1984).

}
\lref\ZamolodchikovXX{
Al.~B.~Zamolodchikov, 
``Conformal Symmetry in Two-Dimensional Space: Recursion Representation of Conformal Block,''
Theor.\ Math.\ Phys.\ {\bf 73}, 103-110 (1987).
}

\lref\ChangJTA{
  C.~M.~Chang, Y.~H.~Lin, S.~H.~Shao, Y.~Wang and X.~Yin,
  ``Little String Amplitudes (and the Unreasonable Effectiveness of 6D SYM),''
[arXiv:1407.7511 [hep-th]].
}

\lref\MaldacenaIUA{
  J.~Maldacena, D.~Simmons-Duffin and A.~Zhiboedov,
  ``Looking for a bulk point,''
JHEP {\bf 1701}, 013 (2017).
[arXiv:1509.03612 [hep-th]].
}

\lref\FroissartUX{
  M.~Froissart,
  ``Asymptotic behavior and subtractions in the Mandelstam representation,''
Phys.\ Rev.\  {\bf 123}, 1053 (1961).
}

\lref\KatzRLA{
  E.~Katz, S.~Sachdev, E.~S.~Sørensen and W.~Witczak-Krempa,
  ``Conformal field theories at nonzero temperature: Operator product expansions, Monte Carlo, and holography,''
Phys.\ Rev.\ B {\bf 90}, no. 24, 245109 (2014).
[arXiv:1409.3841 [cond-mat.str-el]].
}

\lref\BanWil{
S.~Banerjee, B.~Wilkerson, 
``Lambert series and q-functions near $q=1$,'' 
[arXiv:1602.01085 [math.NT]].
}

\lref\KusukiNMS{
  Y.~Kusuki,
  ``Large $c$ Virasoro Blocks from Monodromy Method beyond Known Limits,''
JHEP {\bf 1808}, 161 (2018).
[arXiv:1806.04352 [hep-th]].
}

\lref\HartmanOAA{
  T.~Hartman, C.~A.~Keller and B.~Stoica,
  ``Universal Spectrum of 2d Conformal Field Theory in the Large c Limit,''
JHEP {\bf 1409}, 118 (2014).
[arXiv:1405.5137 [hep-th]].
}

\lref\KrausNWO{
  P.~Kraus and A.~Maloney,
  ``A cardy formula for three-point coefficients or how the black hole got its spots,''
JHEP {\bf 1705}, 160 (2017).
[arXiv:1608.03284 [hep-th]].
}

\lref\HikidaKHG{
  Y.~Hikida, Y.~Kusuki and T.~Takayanagi,
  ``Eigenstate thermalization hypothesis and modular invariance of two-dimensional conformal field theories,''
Phys.\ Rev.\ D {\bf 98}, no. 2, 026003 (2018).
[arXiv:1804.09658 [hep-th]].
}

\lref\HellermanBU{
  S.~Hellerman,
  ``A Universal Inequality for CFT and Quantum Gravity,''
JHEP {\bf 1108}, 130 (2011).
[arXiv:0902.2790 [hep-th]].
}

\lref\CarlipNV{
  S.~Carlip,
  ``Logarithmic corrections to black hole entropy from the Cardy formula,''
Class.\ Quant.\ Grav.\  {\bf 17}, 4175 (2000).
[gr-qc/0005017].
}

\lref\CappelliHF{
  A.~Cappelli, C.~Itzykson and J.~B.~Zuber,
  ``Modular Invariant Partition Functions in Two-Dimensions,''
Nucl.\ Phys.\ B {\bf 280}, 445 (1987)..
}

\lref\FriedanCBA{
  D.~Friedan and C.~A.~Keller,
  ``Constraints on 2d CFT partition functions,''
JHEP {\bf 1310}, 180 (2013).
[arXiv:1307.6562 [hep-th]].
}

\lref\CollierCLS{
  S.~Collier, Y.~H.~Lin and X.~Yin,
  ``Modular Bootstrap Revisited,''
JHEP {\bf 1809}, 061 (2018).
[arXiv:1608.06241 [hep-th]].
}

\lref\KrausNWO{
  P.~Kraus and A.~Maloney,
  ``A cardy formula for three-point coefficients or how the black hole got its spots,''
JHEP {\bf 1705}, 160 (2017).
[arXiv:1608.03284 [hep-th]].
}

\lref\CardyQHL{
  J.~Cardy, A.~Maloney and H.~Maxfield,
  ``A new handle on three-point coefficients: OPE asymptotics from genus two modular invariance,''
JHEP {\bf 1710}, 136 (2017).
[arXiv:1705.05855 [hep-th]].
}

\lref\ChoFZO{
  M.~Cho, S.~Collier and X.~Yin,
  ``Genus Two Modular Bootstrap,''
[arXiv:1705.05865 [hep-th]].
}

\lref\KrausPAX{
  P.~Kraus and A.~Sivaramakrishnan,
  ``Light-state Dominance from the Conformal Bootstrap,''
[arXiv:1812.02226 [hep-th]].
}

\lref\BenjaminFHE{
  N.~Benjamin, E.~Dyer, A.~L.~Fitzpatrick and S.~Kachru,
  ``Universal Bounds on Charged States in 2d CFT and 3d Gravity,''
JHEP {\bf 1608}, 041 (2016).
[arXiv:1603.09745 [hep-th]].
}

\lref\WittenKT{
  E.~Witten,
  ``Three-Dimensional Gravity Revisited,''
[arXiv:0706.3359 [hep-th]].
}

\lref\BenjaminKRE{
  N.~Benjamin, E.~Dyer, A.~L.~Fitzpatrick and Y.~Xin,
  ``The Most Irrational Rational Theories,''
[arXiv:1812.07579 [hep-th]].
}

\lref\MaloneyUD{
  A.~Maloney and E.~Witten,
  ``Quantum Gravity Partition Functions in Three Dimensions,''
JHEP {\bf 1002}, 029 (2010).
[arXiv:0712.0155 [hep-th]].
}

\lref\PonsotUF{
  B.~Ponsot and J.~Teschner,
  ``Liouville bootstrap via harmonic analysis on a noncompact quantum group,''
[hep-th/9911110].
}

\lref\PonsotMT{
  B.~Ponsot and J.~Teschner,
  ``Clebsch-Gordan and Racah-Wigner coefficients for a continuous series of representations of U(q)(sl(2,R)),''
Commun.\ Math.\ Phys.\  {\bf 224}, 613 (2001).
[math/0007097 [math-qa]].
}

\lref\KusukiWPA{
  Y.~Kusuki,
  ``Light Cone Bootstrap in General 2D CFTs and Entanglement from Light Cone Singularity,''
JHEP {\bf 1901}, 025 (2019).
[arXiv:1810.01335 [hep-th]].
}

\lref\CollierEXN{
  S.~Collier, Y.~Gobeil, H.~Maxfield and E.~Perlmutter,
  ``Quantum Regge Trajectories and the Virasoro Analytic Bootstrap,''
[arXiv:1811.05710 [hep-th]].
}

\lref\DetournayPC{
  S.~Detournay, T.~Hartman and D.~M.~Hofman,
  ``Warped Conformal Field Theory,''
Phys.\ Rev.\ D {\bf 86}, 124018 (2012).
[arXiv:1210.0539 [hep-th]].
}

\lref\SongTXA{
  W.~Song and J.~Xu,
  ``Structure Constants from Modularity in Warped CFT,''
[arXiv:1903.01346 [hep-th]].
}

\lref\ShifmanJV{
  M.~A.~Shifman,
  ``Quark hadron duality,''
[hep-ph/0009131].
}

\lref\YndurainMZ{
  F.~J.~Yndurain,
  ``Absolute bound on cross-sections at all energies and without unknown constants,''
Phys.\ Lett.\  {\bf 31B}, 368 (1970).
}

\lref\LashkariVGJ{
  N.~Lashkari, A.~Dymarsky and H.~Liu,
  ``Eigenstate Thermalization Hypothesis in Conformal Field Theory,''
J.\ Stat.\ Mech.\  {\bf 1803}, no. 3, 033101 (2018).
[arXiv:1610.00302 [hep-th]].
}

\lref\BelinYLL{
  A.~Belin, J.~de Boer, J.~Kruthoff, B.~Michel, E.~Shaghoulian and M.~Shyani,
  ``Universality of sparse $d > 2$ conformal field theory at large $N$,''
JHEP {\bf 1703}, 067 (2017).
[arXiv:1610.06186 [hep-th]].
}

\lref\MazacQEV{
  D.~Mazac,
  ``Analytic bounds and emergence of AdS$_{2}$ physics from the conformal bootstrap,''
JHEP {\bf 1704}, 146 (2017).
[arXiv:1611.10060 [hep-th]].
}

\lref\MazacYCV{
  D.~Mazac and M.~F.~Paulos,
  ``The analytic functional bootstrap. Part II. Natural bases for the crossing equation,''
JHEP {\bf 1902}, 163 (2019).
[arXiv:1811.10646 [hep-th]].
}

\lref\MazacMDX{
  D.~Mazac and M.~F.~Paulos,
  ``The analytic functional bootstrap. Part I: 1D CFTs and 2D S-matrices,''
JHEP {\bf 1902}, 162 (2019).
[arXiv:1803.10233 [hep-th]].
}

\lref\FrenkelXZ{
  I.~Frenkel, J.~Lepowsky and A.~Meurman,
  ``Vertex Operator Algebras And The Monster,''
BOSTON, USA: ACADEMIC (1988) 508 P. (PURE AND APPLIED MATHEMATICS, 134).
}

\lref\SenDW{
  A.~Sen,
  ``Logarithmic Corrections to Schwarzschild and Other Non-extremal Black Hole Entropy in Different Dimensions,''
JHEP {\bf 1304}, 156 (2013).
[arXiv:1205.0971 [hep-th]].
}

\lref\StromingerSH{
  A.~Strominger and C.~Vafa,
  ``Microscopic origin of the Bekenstein-Hawking entropy,''
Phys.\ Lett.\ B {\bf 379}, 99 (1996).
[hep-th/9601029].
}

\lref\SenQY{
  A.~Sen,
  ``Black Hole Entropy Function, Attractors and Precision Counting of Microstates,''
Gen.\ Rel.\ Grav.\  {\bf 40}, 2249 (2008).
[arXiv:0708.1270 [hep-th]].
}

\lref\AfkhamiJeddiZCI{
  N.~Afkhami-Jeddi, T.~Hartman and A.~Tajdini,
  ``Fast Conformal Bootstrap and Constraints on 3d Gravity,''
[arXiv:1903.06272 [hep-th]].
}

\lref\FriedanCBA{
  D.~Friedan and C.~A.~Keller,
  ``Constraints on 2d CFT partition functions,''
JHEP {\bf 1310}, 180 (2013).
[arXiv:1307.6562 [hep-th]].
}

\lref\MarolfLDL{
  D.~Marolf,
  ``Microcanonical Path Integrals and the Holography of small Black Hole Interiors,''
JHEP {\bf 1809}, 114 (2018).
[arXiv:1808.00394 [hep-th]].
}

\lref\deLangeMRI{
  P.~De Lange, A.~Maloney and E.~Verlinde,
  ``Monstrous Product CFTs in the Grand Canonical Ensemble,''
[arXiv:1807.06200 [hep-th]].
}

\lref\BrownBQ{
  J.~D.~Brown and J.~W.~York, Jr.,
  ``The Microcanonical functional integral. 1. The Gravitational field,''
Phys.\ Rev.\ D {\bf 47}, 1420 (1993).
[gr-qc/9209014].
}

\lref\MazacYCV{
  D.~Mazac and M.~F.~Paulos,
JHEP {\bf 1902}, 163 (2019).
[arXiv:1811.10646 [hep-th]].
}

\lref\SenVZ{
  A.~Sen,
  ``Arithmetic of Quantum Entropy Function,''
JHEP {\bf 0908}, 068 (2009).
[arXiv:0903.1477 [hep-th]].
}

\lref\MandalCJ{
  I.~Mandal and A.~Sen,
  ``Black Hole Microstate Counting and its Macroscopic Counterpart,''
Nucl.\ Phys.\ Proc.\ Suppl.\  {\bf 216}, 147 (2011), [Class.\ Quant.\ Grav.\  {\bf 27}, 214003 (2010)].
[arXiv:1008.3801 [hep-th]].
}

\hfil \hfil \hfil \hfil \hfil \hfil \hfil CERN-TH-2019-043

\Title{
\vbox{\baselineskip6pt
}}
{\vbox{
\centerline{Modular Invariance, Tauberian Theorems,}
\centerline{and Microcanonical Entropy}
}}

\bigskip
\centerline{Baur Mukhametzhanov\footnote{$^\dagger$}{baurinho@gmail.com}$^{\Delta}$ 
and Alexander Zhiboedov$^{\beta}$
}
\bigskip
\centerline{\it $^\Delta$ Department of Physics, Harvard University, Cambridge, MA 02138, USA}
\centerline{\it $^\beta$  CERN, Theoretical Physics Department, 1211 Geneva 23, Switzerland
}

\vskip .2in

\noindent

We analyze modular invariance drawing inspiration from tauberian theorems.  Given a modular invariant partition function with a positive spectral density, we derive lower and upper bounds on the number of operators within a given energy interval. They are most revealing at high energies. In this limit we rigorously derive the Cardy formula for the microcanonical entropy together with optimal error estimates for various widths of the averaging energy shell. We identify a new universal contribution to the microcanonical entropy controlled by the central charge and the width of the shell. We derive an upper bound on the spacings between Virasoro primaries. Analogous results are obtained in holographic 2d CFTs.  We also study partition functions with a UV cutoff. Control over error estimates allows us to probe operators beyond the unity in the modularity condition. We check our results in the 2d Ising model and the Monster CFT and find perfect agreement.

\Date{April 2019}

\listtoc\writetoc
\vskip .1in \noindent

\break

\newsec{Introduction}

High energy estimates on various physical quantities are commonly stated locally even though they are only true on average. A few famous examples are: the Froissart bound on the growth of the cross section \refs{\FroissartUX, \YndurainMZ}, high frequency expansion of conductivity at finite temperature \refs{\KatzRLA,\CaronHuotNS}, high energy asymptotic of the electromagnetic current spectral density in the context of the so-called quark-hadron duality \ShifmanJV, and finally the Cardy formula for two-dimensional CFTs \CardyIE.  The latter is particularly interesting because of its importance for the problem of the black hole microstate counting \refs{\StromingerSH\StromingerEQ\SenQY-\SenDW}.  It is then a natural question to ask: how do these estimates depend on the details of the averaging?

Let us review the standard derivation of the Cardy formula. We consider a thermal partition function $Z(\beta)$ of a unitary 2d CFT on a Euclidean torus. The partition function is modular invariant $Z(\beta) = Z({4 \pi^2 \over \beta})$. This implies that the high-temperature limit $\beta \to 0$ of the partition function is captured by the contribution of the vacuum in the dual channel $Z(\beta) \sim e^{{\pi^2 c \over 3 \beta}}$, where $c$ is the central charge. Using the standard thermodynamic formula $S(\beta) = (1- \beta \partial_\beta) \log Z$ we can compute the entropy $S(\beta)$ at high temperatures
\eqn\CardyZ{
S(\beta) =   {2 \pi^2 c \over 3 \beta}+ ... , ~~~ \beta \to 0 \ .
}
In the $\beta \to 0 $ limit the energy of the system $\la \Delta \ra = - \partial_\beta \log Z = {\pi^2 c \over 3\beta^2} + ...$ goes to infinity. The $\Delta \to \infty$ limit being the thermodynamic limit, see e.g. \LashkariVGJ , one obtains from the usual thermodynamic arguments that  the extensive part of the entropy, which is given by \CardyZ, also correctly captures the leading behavior of the microcanonical entropy $S_{\delta}(\Delta)$ defined by
\eqn\microcan{\eqalign{
S_{\delta}(\Delta) &\equiv \log  \int_{\Delta- \delta}^{\Delta+ \delta} d \Delta' \rho(\Delta') , \cr
\rho(\Delta) &\equiv \sum_{\cal O} \delta(\Delta - \Delta_{\cal O}) ,
}}
as soon as $\delta$ is large enough to include many energy levels. That is if we express the temperature as a function of the average energy $\beta = \pi \sqrt{c/3 \la\Delta \ra}$ and plug it in \CardyZ\ we arrive at the famous Cardy formula for the micronaconical entropy
\eqn\Cardy{
S_\delta(\Delta) = 2 \pi \sqrt{ {c \Delta \over 3} } + ... , ~~~ \Delta \to \infty .
}
In discussions and applications of the Cardy formula the averaging width parameter $\delta$ is usually kept implicit. Moreover, the rigorous transition from \CardyZ\ to \Cardy\ requires some extra work which is usually left to the reader.

Indeed, the spectral density $\rho(\Delta)$ is related to the partition function $Z(\beta)$ by the inverse Laplace transform. It is sometimes argued that this Laplace transform can be evaluated by a saddle point approximation from which the statement about $\rho(\Delta)$ and therefore $S_{\delta}(\Delta)$ can be made. A more accurate description of this procedure would be to say that one can easily find the crossing kernel of the vacuum contribution $e^{{\pi^2 c \over 3 \beta}}$, or, in other words, a spectral density $\rho_0(\Delta)$ that correctly reproduces the vacuum in the dual channel. The question then stays: what is the precise relation between the naive spectral density $\rho_0(\Delta)$ and the actual physical density $\rho(\Delta)$? This relation cannot be too literal. Indeed, the former is a smooth function of $\Delta$, whereas the latter is a sum of delta-functions. Once again the physical intuition is that they are related on average, but establishing this rigorously is a nontrivial task.  The issue of making the argument precise becomes even more important if one considers ``finite size'' or ${1 \over \Delta}$ corrections to the Cardy formula. The purpose of this paper is to close this gap in the usual discussions of the Cardy formula and to develop further techniques that allow us to study ${1 \over \Delta}$ corrections to it.

The physical question of going from the finite temperature partition function to the microcanonical entropy can be addressed in a mathematically rigorous way using the methods of tauberian theory \Korevaar, as explained in \refs{\PappadopuloJK,\QiaoXIF}. From the conformal/modular bootstrap point of view tauberian theory provides a natural set of linear functionals with which we act on the crossing/modularity condition to derive optimal estimates on $S_{\delta}(\Delta)$ or other spectral density averages.

 As further noticed in \MukhametzhanovZJA\ the optimal error estimates can be obtained using the so-called complex tauberian theorems, which exploit the fact that physical quantities of interest are very often analytic functions in a complex domain. This is indeed the case for the modularity condition of 2d CFTs. In this note we apply methods of tauberian theory to modular invariance in 2d CFTs and rigorously derive the Cardy formula and corrections to it, where we explicitly keep track of the dependence on $\delta$.
Furthermore, combining these ideas with bounds on the the partition function of Hartman, Keller and Stoica (HKS) \HartmanOAA\ we find lower and upper bounds on the number of operators within a given window of finite conformal dimensions $(\Delta - \delta, \Delta + \delta)$. Though true at finite $\Delta$, they are most revealing in the limit $\Delta \to \infty$.

\subsec{Review of the Results}

We consider a modular invariant partition function with zero angular potential and positive spectral density. We derive a set of rigorous results about $S_{\delta}(\Delta)$ \microcan. These concern either all operators present in the theory, or only Virasoro primaries in CFTs with $c>1$. 

$\bullet$ Let us first discuss densities of all operators, both primaries and descendants. We derive a rigorous asymptotic for the microcanonical entropy 
\eqn\IntroMicroCardy{\eqalign{
S_\delta(\Delta) &= \log \int_{\Delta - \delta}^{\Delta + \delta} d \Delta'\rho(\Delta') = 2\pi\sqrt{c\Delta \over 3} +  {1\over 4}\log \left( c \delta^4\over 3  \Delta^3  \right)+  s(\delta, \Delta) \ , ~~~ \Delta \to \infty \ , 
}}
where depending on the size of the averaging energy shell $\delta$ we show that\foot{By $a \sim b$ we mean $\lim a/b = const \neq 0$ in the corresponding limit.}
\eqn\sboundpm{\eqalign{
\delta &\sim \Delta^\alpha  : \quad s(\delta , \Delta) = 
   \log \left(  {\sinh \left( \pi \sqrt{c\over 3} {\delta \over \sqrt \Delta} \right) \over   \pi \sqrt{c\over 3} {\delta \over \sqrt \Delta}     } \right) +  O\left( \Delta^{-\alpha} \right)\ , \quad 0 < \alpha \leq {1 \over 2} \ , \cr
\delta &= O(1) : \quad s_-(\delta) \leq s(\delta, \Delta) \leq s_+(\delta)  \ , ~~~ \delta > \delta_{gap} = {\sqrt{3} \over \pi} \approx 0.55 \ .
}}

\ifig\sbound{On this figure we plot the upper (yellow) and lower (blue) bounds $s_{\pm}(\delta)$ on $s(\delta,\Delta)$. The vertical line is $\delta = \delta_{gap} = {\sqrt{3 \over \pi}}$, below which we do not have a lower bound. The divergence of $s_+(\delta)$ at $\delta = 0$ is spurious and is cancelled by $\log \delta$ in \IntroMicroCardy. }{\epsfxsize3.5in\epsfbox{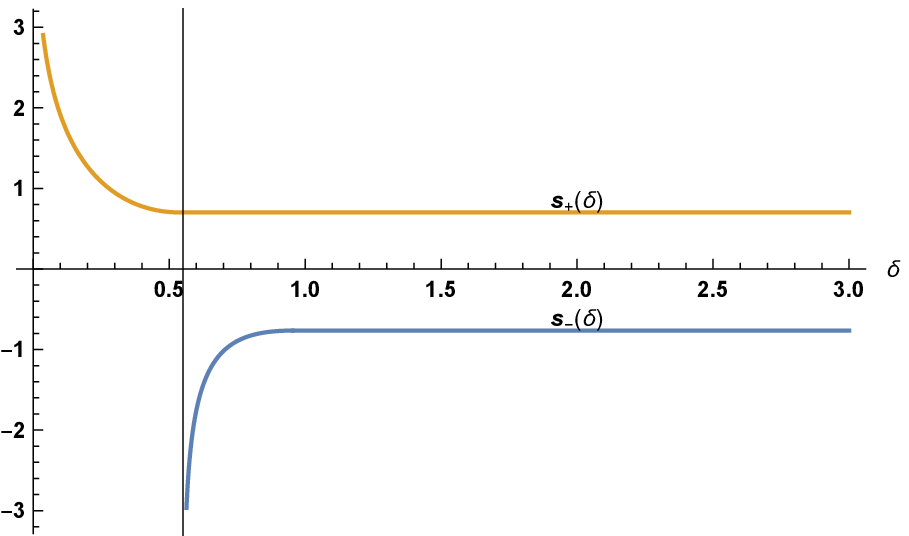}}

The first two terms in the RHS of \IntroMicroCardy\ are the Cardy formula \Cardy\ and the leading log correction to it discussed for example in \refs{\CarlipNV,\SenDW}. The results for $s(\delta, \Delta)$ are new to the best of our knowledge. In particular, we see that for $\delta \sim \Delta^\alpha$ there is yet another universal correction\foot{It dominates over the error term $O\left( \Delta^{-\alpha} \right)$ only for $\alpha>{1 \over 3}$.} to the microcanonical entropy that is controlled by the central charge $c$ and the width of the energy shell $\delta$ and given by the first line in \sboundpm. Note that for any $\alpha>0$ the error decays at large $\Delta$ and the non-decaying contribution to the entropy is fully captured by the ${1\over 4}\log \left( c \delta^4\over 3  \Delta^3  \right)$ term. For $\delta = O(1)$ the functions $s_\pm(\delta)$ are plotted on \sbound. In particular, the lower bound diverges logarithmically as we approach $\delta \to \delta_{gap}$. The interpretation of this is that the asymptotic \IntroMicroCardy\ is only applicable for $\delta > \delta_{gap}$ for which the leading behavior of the microcanonical entropy $S_{\delta}(\Delta)$ takes the form \IntroMicroCardy. Note that the lower bound implies that there have to be operators in an energy shell of the size $\delta > \delta_{gap}$. 

For $\delta < \delta_{gap}$ we can only prove an upper bound on the microcanonical entropy which is given by \IntroMicroCardy\ and $s_+(\delta)$ in \sboundpm .\foot{The precise form of the bounding curves can be found in section 4. } For a fixed $\delta = O(1)$ the function $s(\delta,\Delta)$ in general is not a constant and can oscillate as we change $\Delta$, but always between the values $s_\pm(\delta)$. In fact, we will explicitly see these oscillations in the 2d Ising model in section 7.

$\bullet$ For CFTs with $c>1$ we can derive analogous formulae for Virasoro primary operators 
\eqn\IntroMicroCardyVir{\eqalign{
S^{{\rm Vir}}_\delta(\Delta) &= \log  \int_{\Delta - \delta}^{\Delta + \delta} d \Delta' \rho^{{\rm Vir}}(\Delta') =2\pi \sqrt{ {c-1 \over 3 } \Delta} - {1\over 4} \log \left( {c-1 \over 48 \delta^4} \Delta \right)  + s^{\rm Vir}(\delta, \Delta),  \cr
\delta &\sim \Delta^\alpha  : \quad s^{\rm Vir}(\delta , \Delta) = 
   \log \left(  {\sinh \left( \pi \sqrt{c-1\over 3} {\delta \over \sqrt \Delta} \right) \over   \pi \sqrt{c-1\over 3} {\delta \over \sqrt \Delta}     } \right) +  O\left( \Delta^{-\alpha} \right) , \quad 0 < \alpha \leq {1 \over 2}  , \cr
\delta &= O(1) : \quad s_-(\delta) \leq s^{\rm Vir}(\delta, \Delta) \leq s_+(\delta)  \ , ~~~ \delta > \delta_{gap} = {\sqrt{3} \over \pi} \approx 0.55 \ .
}}
where $\Delta \to \infty$ and $s_\pm(\delta)$ are the same as in \sboundpm. For finite width energy shells, or $\alpha = 0$, we can again write the lower and upper bounds on the entropy as in \sboundpm , as soon as $\delta > \delta_{gap}$. A simple consequence of this result is an existence of maximal sparseness of Virasoro primaries. In other words, it follows that

$\bullet$ At large scaling dimensions $\Delta$ the spacings between Virasoro primary operators in CFTs with $c>1$ cannot be larger than $2\delta_{gap} = {2\sqrt{3} \over \pi} \approx 1.1$.

\vskip .1in

\noindent This bound is not necessarily optimal. Nevertheless, it is close to the optimal since there are many examples of theories with the spacings equal to $1$.\foot{A famous example is the monster CFT \refs{\FrenkelXZ,\WittenKT}. The monster CFT is chiral with $(c_L, c_R)=(24,0)$. However, for zero angular potential its partition function can be interpreted as the one of a non-chiral theory with $(c_L, c_R)=(12,12)$, see e.g. \AfkhamiJeddiZCI. It, therefore, satisfies the modularity constraint imposed in this paper.}

$\bullet$ We derive an asymptotic of the microcanonical entropy in holographic 2d CFTs in the limit $c\to \infty$ with $\Delta/c$ - fixed and $\Delta > {c\over 6}$
\eqn\IntroMicroLargeC{\eqalign{
S_\delta(\Delta) &= \log \int_{\Delta - \delta}^{\Delta + \delta} d \Delta'\rho(\Delta') = 2\pi\sqrt{{c\over 3}\left( \Delta + \delta - {c\over 12} \right)} -  {1\over 2}\log c+O(1) \ , ~~~ c \to \infty \ , 
}}
where $\delta \sim c^\alpha, 0 \leq \alpha < 1$ and $ \delta > \delta_{gap}$. This relies on the sparseness condition of Hartman, Keller and Stoica (HKS) \HartmanOAA\ and extends their result\foot{See appendix A in their paper.} for the microcanonical entropy which is \IntroMicroLargeC\ with an extra constraint ${1 \over 2} <\alpha < 1$.

As we will explain later on, an important ingredient in the derivation of $s_\pm(\delta)$ and $\delta_{gap}$ relies on the existence of functions $\phi_\pm(\Delta')$ with the following properties: 

1) $\phi_+(\Delta')$ and $\phi_-(\Delta')$ bound the indicator function of the interval $(\Delta - \delta, \Delta + \delta)$ from above and below respectively; 

2) Their Fourier transform has a bounded support.  

\noindent We make an explicit choice of such functions to arrive at the particular value of $\delta_{gap}$ and the bounding curves in $s_\pm(\delta)$. Nevertheless, the method is completely general and we leave open the question of finding the functions $\phi_\pm(\Delta')$ giving optimal bounds.

$\bullet$ Above we stated our results at asymptotically high energies. They follow from more general bounds on the number of operators at finite $\Delta,c$, that we derive in section 4. Specifically, given the data about operators $\Delta \leq c/12$ we derive rigorous upper and lower bounds on the number of operators in a given window of scaling dimensions. We emphasize that all parameters can be kept finite. In particular, these bounds can be easily implemented numerically. For example, we can derive numerical bounds on the gap above the vacuum, though these turn out to be weaker than \CollierCLS, \AfkhamiJeddiZCI. On the other hand we can also bound a number of operators in any window of scaling dimensions at any $\Delta$ above the first excited state as well.\foot{Analogous bounds for the spectral density weighted by the squares of OPE coefficients in 1d CFTs were recently derived in \MazacYCV.}

$\bullet$ We consider partition functions with a UV cutoff. We start by proving a generalized Ingham's theorem:
\vskip .1in
{\bf Theorem:} Consider a positive spectral density $\rho(\Delta)$, such that the partition function $Z(\beta) = \int_0^{\infty} d \Delta e^{- \beta (\Delta - {c \over 12})} \rho(\Delta)$ is modular invariant $Z(\beta) = Z({4 \pi^2 \over \beta})$. Moreover, suppose that $Z(\beta) = e^{{\pi^2 c \over 3 \beta}} \left( 1 + O(e^{- {\Delta_1 \over \beta}}) \right)$ with $\Delta_1 >0$, when $|\beta| \to 0$ for ${\rm Re}[\beta] >0$. 
Then the integrated spectral density satisfies
\eqn\introTrueCardy{
F_\rho(\Delta) \equiv \int_0^\Delta d\Delta' ~ \rho(\Delta') ={1\over 2\pi} \left( 3\over c\Delta \right)^{1/4} e^{2\pi \sqrt{{c\over 3} \Delta}} \left[ 1 + O(\Delta^{-1/2}) \right], \qquad \Delta \to \infty \ .
}
\vskip .1in

The RHS of \introTrueCardy\ comes from the unit operator in the dual modular channel, which dominates the partition function at high temperatures. The average of the physical density of states in the LHS side of \introTrueCardy\ is a discontinuous ``staircase-like" function. It is approximated by a smooth function in the RHS of \introTrueCardy\ with a bounded error term. The discontinuities of the LHS of \introTrueCardy\ are hidden in the non-universal\foot{Everywhere in this paper by ``non-universal terms'' we mean the terms that are not controlled by light operators in the dual channel.} error term in the RHS. In particular, it does not make sense to write further smooth power suppressed terms in the RHS of \introTrueCardy. We will see it explicitly in the example of 2d Ising model that the error term is a highly oscillating function and cannot be approximated by a smooth function. This example will also demonstrate that the error estimate is optimal.

The asymptotic \IntroMicroCardy, \sboundpm\ of the microcanonical entropy for energy shells $\delta \sim \Delta^\alpha, 0<\alpha <1$ follows directly from \introTrueCardy. Further, using this theorem we derive a bound on the cutoff partition function at finite temperature\foot{And a similar bound for $\beta < \pi \sqrt{c\over 3\Delta}$.}
\eqn\Zcutoff{\eqalign{
\int_0^{\Delta} d\Delta' ~  \rho(\Delta') e^{-\beta (\Delta'-c/12)} &= \int_0^{\Delta} d\Delta' ~  \rho_0(\Delta') e^{-\beta (\Delta'-c/12)}    \cr 
&+Z \left({4 \pi^2 \over \beta}\right) - e^{\pi^2 c \over 3 \beta} +O\left(\Delta^{-3/4} e^{2\pi \sqrt{{c\over 3} \Delta} - \beta \Delta} \right), \quad \beta > \pi \sqrt{c\over 3 \Delta} \ , 
}}
where $\rho_0$ is the vacuum crossing kernel defined below. Depending on the temperature some operators in the dual channel in the RHS of \Zcutoff\ dominate over the error term and therefore are captured by the cutoff partition function in the LHS.

\subsec{Related Works}

The averaging procedure \introTrueCardy\ was first pointed out in the context of CFTs in \PappadopuloJK. In the mathematical literature the asymptotic \introTrueCardy\ without the error estimate is known as Ingham's tauberian theorem for large Laplace transform \Ingham. For a nice exposition of this result see \Korevaar, Section IV.21. The relevance of Ingham's theorem for Cardy formula was also emphasized in \DasVEJ, Appendix C.  We give a derivation of \introTrueCardy, which is different from the original proof \Ingham. The novelty of \introTrueCardy\ is the error estimate which is absent in the Ingham's theorem. In the proof we use the methods of \SubhankulovR, Section 2.3,  extensively discussed in \MukhametzhanovZJA. In particular, the error estimate allows us to access subleading operators in the cutoff partition function \Zcutoff.

\newsec{Setup}

Consider a unitary 2d CFT on a torus with the modular parameter $\tau = {1\over 2\pi} (\theta + i \beta)$ and the coordinate on the torus $z = {1\over 2\pi} (\phi + i t_E)$ with standard identifications $z \sim z+1 \sim z + \tau$. In these conventions the spatial circle $\phi$ has length $2\pi$ and the Euclidean time circle $t_E$ has length $\beta$. The partition function 
\eqn\pf{
Z(\tau, \b \tau) = \Tr ~ q^{L_0 - c/ 24} \bar q^{\bar L_0 - c/ 24}, \qquad q = e^{2\pi i \tau} 
}
is invariant under the modular transformation $\tau \to -1/\tau$. In what follows we restrict to zero angular potential $\theta = 0$ so that $q = e^{-\beta}$.  However, we consider complex $\beta$ with ${\rm Re}[\beta]>0$. This is possible due to unitarity.\foot{Unitarity implies that degeneracies of operators are positive. Therefore, for complex $\beta$ the trace in \pf\ converges even better than for real $\beta$ and, hence, finite.} In this case the modular invariance is expressed by
\eqn\modinv{
Z(\beta) = Z\left({4\pi^2 \over \beta}\right), ~~~ {\rm Re}[\beta]>0 ,
}
or, equivalently,
\eqn\modinvv{
\int_0^\infty d\Delta ~ \rho(\Delta) e^{-\beta (\Delta - c/12) } = \int_0^\infty d\Delta ~ \rho(\Delta) e^{-{4 \pi^2 \over \beta} (\Delta - c/12) } ,
}
where the density of states is defined by 
\eqn\spec{
\rho(\Delta) = \sum_{\cal O} \delta(\Delta - \Delta_{\cal O})
}
and the sum is over all operators in the theory, both primaries and descendants. We will be interested in exploring consequences of \modinv.
\foot{For some rational CFTs the solutions to \modinv\ were classified \CappelliHF.} 

In the high-temperature limit $|\beta| \to 0$ the RHS of \modinv\ is dominated by the unit operator 
\eqn\highT{
Z(\beta) = e^{\pi^2 c \over 3 \beta} \left[ 1 + O(e^{-{4\pi^2 \over \beta} \Delta_1} ) \right]  \ ,
}
where $\Delta_1$ is the first operator above the vacuum.

To write the asymptotic of spectral density it will convenient to introduce a ``naive'' spectral density $\rho_0(\Delta)$ which correctly reproduces the contribution of the vacuum in the partition function.
The correct expression takes the form
\eqn\naiveCardy{\eqalign{
\rho_{0}(\Delta) = &\pi \sqrt{ c \over 3} ~{ I_1\left( 2\pi \sqrt{ {c\over 3} \left( \Delta- {c\over 12} \right) } \right) 
\over  \sqrt{  \Delta - {c\over 12} } } \theta(\Delta - c/12)  + \delta(\Delta - c / 12)\cr 
 =&  \left( c\over 48 \Delta^3\right)^{1/4} e^{2\pi \sqrt{c\Delta/3}}  \left[ 1 + O(\Delta^{-1/2})\right] \theta (\Delta - c/12) + \delta(\Delta - c/12)
}}
where $\theta(x)$ is the Heaviside step function. This, of course, cannot be literally an approximation of the physical density of states \spec, as the latter is a sum of delta functions. The index $``0"$ in the LHS of \naiveCardy\ is reminding us of that. Nevertheless, the Laplace transform of \naiveCardy\ coincides with the unit operator contribution into the partition function
\eqn\Lofnaive{
\int_0^\infty d\Delta ~ \rho_0 (\Delta) e^{-\beta (\Delta - c/12)} = e^{\pi^2 c \over 3 \beta}  .
}
The function $\rho_0(\Delta)$ can be naturally called ``crossing kernel'' in analogy with \CollierEXN.

\newsec{HKS Bound on Heavy Operators}

\noindent An important result for obtaining bounds on the spectral density at finite $\Delta$ will be the bound of Hartman, Keller, Stoica (HKS bound) \HartmanOAA\ on the contribution of heavy operators into the partition function. We review its derivation in this section.

We split the partition function as
\eqn\partsplit{\eqalign{
&Z(\beta) = Z_L(\beta) + Z_H(\beta), \cr 
&Z_L(\beta) = \sum_{\Delta < \Delta_H} e^{-\beta (\Delta - c/12)}, \qquad Z_H(\beta) = \sum_{\Delta \geq \Delta_H} e^{-\beta (\Delta - c/12)} \ .
}}
Modular invariance states that 
\eqn\modular{
Z_L + Z_H = Z_L' + Z_H' \ ,
}
where by primes we denote the dual channel $\beta' = {4\pi^2 \over \beta} $. Suppose $\beta \geq 2\pi$. We would like to estimate $Z_H$
\eqn\ZHest{\eqalign{
Z_H =& \sum_{\Delta \geq \Delta_H} e^{-(\beta - \beta') (\Delta - c/12)}  e^{-\beta' (\Delta - c/12)} \leq 
e^{-(\beta - \beta') (\Delta_H - c/12)} Z_H' \cr 
=& e^{-(\beta - \beta') (\Delta_H - c/12)} (Z_H + Z_L - Z_L') \ .
}}
Now if $\Delta_H > c/12$ then \ZHest\ implies an upper bound on $Z_H$
\eqn\HKSa{
Z_H \leq   e^{-(\beta - \beta') (\Delta_H - c/12)}{ Z_L - Z_L'   \over 1 - e^{-(\beta - \beta') (\Delta_H - c/12)}} , \qquad \beta \geq 2\pi \ .
}
This also implies a bound on $Z_H'$ via modular invariance
\eqn\ZHprest{
Z_H' = Z_H + Z_L - Z_L' \leq {  Z_L - Z_L'  \over 1 - e^{-(\beta - \beta') (\Delta_H - c/12)}}, \qquad \beta \geq 2\pi \ .
}
Exchanging $\beta$ and $\beta'$ in \ZHprest\ we can turn it into a bound at high temperatures
\eqn\HKSb{
Z_H \leq  { Z_L' - Z_L  \over 1 - e^{-(\beta' - \beta) (\Delta_H - c/12)}} , \qquad \beta \leq 2\pi \ .
}
Depending on the temperature the bound on the heavy operators is either \HKSa\ or \HKSb. Everywhere we assume that $\Delta_H > c/12$.

Finally, \HKSa, \HKSb\ lead to bounds on the full partition function
\eqn\HKSc{\eqalign{
Z & \leq  { 1  \over 1 - e^{-(\beta- \beta') (\Delta_H - c/12)}}  \left[ Z_L - e^{-(\beta- \beta') (\Delta_H - c/12)} Z_L' \right], \qquad \beta \geq 2\pi  \ ,\cr 
Z & \leq  { 1  \over 1 - e^{-(\beta' - \beta) (\Delta_H - c/12)}}  \left[ Z_L' - e^{-(\beta' - \beta) (\Delta_H - c/12)} Z_L \right], \qquad \beta \leq 2\pi \ .
}}
Note that the bounds \HKSa, \HKSb\ stay finite if we take $\beta \to 2\pi$. Indeed, $Z_L - Z_L' $ is zero and cancels the zero of the denominator. Whereas $\Delta_H$ is strictly above the BTZ threshold $c\over 12$.

\newsec{Local Bound on the Number of Operators}

We can use modular invariance together with the HKS bound to derive a local bound on the density of operators. To that end let us consider two functions $\phi_{\pm}(\Delta)$ such that
\eqn\functions{
\phi_{-} (\Delta') \leq \theta_{[\Delta - \delta, \Delta+\delta]}(\Delta') \leq \phi_+(\Delta')  \ ,
}
where $\theta_{[\Delta - \delta, \Delta+\delta]} (\Delta') =\theta \left(\Delta' \in [\Delta - \delta, \Delta+\delta] \right)$.

We can multiply this inequality by $e^{-\beta \Delta'}$ and use $e^{\beta(\Delta - \delta) }e^{- \beta \Delta'} \theta_{[\Delta - \delta, \Delta+\delta]}  \leq \theta_{[\Delta - \delta, \Delta+\delta]} \leq e^{\beta(\Delta+\delta)} e^{- \beta \Delta'}\theta_{[\Delta - \delta, \Delta+\delta]}$ to write
\eqn\functionsB{
e^{\beta (\Delta - \delta)} e^{- \beta \Delta'}\phi_{-} (\Delta')  \leq \theta_{[\Delta - \delta, \Delta+\delta]}(\Delta') \leq   e^{\beta (\Delta + \delta)} e^{- \beta \Delta'}\phi_+(\Delta')  .
}

Integrating both sides of \functionsB\ with the spectral density $\int_0^{\infty} d F(\Delta')$ we finally obtain an estimate
\eqn\boundsA{
e^{\beta (\Delta - \delta)} \int_0^{\infty} d F(\Delta') e^{-\beta \Delta'} \phi_-(\Delta')   \leq \int_{\Delta - \delta}^{\Delta + \delta} dF(\Delta') \leq e^{\beta (\Delta + \delta)}  \int_0^{\infty} d F(\Delta') e^{-\beta \Delta'} \phi_+(\Delta') .
}
In the inequality above $\beta$ and $\delta$ are free parameters. We will fix $\beta$ below by making the bound optimal.

Next the idea is to do the Fourier transform $\phi_{\pm}(\Delta) = \int_{-\infty}^{ \infty} d t ~\widehat \phi_{\pm}(t) e^{- i \Delta t}$ which turns \boundsA\ into a bound in terms of the partition function
\eqn\boundsB{
 e^{\beta (\Delta - \delta)} \int_{-\infty}^{ \infty} d t ~ \widehat \phi_{-}(t) {\cal L}_{\rho}(\beta+ i t) \leq \int_{\Delta - \delta}^{\Delta + \delta} dF(\Delta') \leq e^{\beta (\Delta + \delta)} \int_{-\infty}^{ \infty} d t  ~ \widehat \phi_{+}(t) {\cal L}_\rho(\beta+ i t) \ ,
}
where we introduced the Laplace transform ${\cal L}$ of a density $\rho$
\eqn\laplace{
{\cal L}_\rho(\beta) \equiv \int_0^\infty d\Delta ~ \rho(\Delta) e^{-\beta \Delta} .
}

As a next step we apply a modular transformation to ${\cal L}(\beta+ i t)$ and separate the contribution of light and heavy operators in the dual channel. We write ${\cal L}(\beta+ i t) = e^{-(\beta+i t) c/12} Z(\beta+i t) = e^{-(\beta+i t) c/12} Z({4 \pi^2 \over \beta+ i t})$ and split $Z = Z_{L} + Z_{H}$. As in \Lofnaive\ we can rewrite $e^{-(\beta+i t) c/12} Z_L({4 \pi^2 \over \beta+ i t}) ={\cal L}_{\rho_0, L}(\beta+ i t) $, where the superscript $\rho_0$ refers to the fact that the Laplace transform is computed with the crossing kernel rather than the density of actual physical operators.\foot{Here it is implied that the crossing kernel $\rho_0$ is not only for the vacuum \Lofnaive, but for all light operators entering $Z_L$. Though in the large $\Delta$ analysis below the vacuum contribution will be dominant.} 
In this way we get
\eqn\boundsBBB{\eqalign{
& e^{\beta (\Delta - \delta)} \left( \int_{-\infty}^{\infty} d t ~ \widehat \phi_{-}(t)  {\cal L}_{\rho_0, L}(\beta+ i t)  -\left| \int_{- \infty}^{ \infty} d t ~ \widehat \phi_{-}(t) e^{-(\beta+i t) {c\over 12}} Z_H({4 \pi^2 \over \beta+ i t}) \right| \right) \cr
&\leq \int_{\Delta - \delta}^{\Delta + \delta} dF(\Delta') \leq \cr
&  e^{\beta (\Delta + \delta)} \left( \int_{- \infty}^{ \infty} d t ~\widehat \phi_{+}(t) {\cal L}_{\rho_0, L}(\beta+ i t)  +  \left|  \int_{- \infty}^{ \infty} d t ~\widehat \phi_{+}(t) e^{-(\beta+i t) {c\over 12}} Z_H({4 \pi^2 \over \beta+ i t}) \right| \right) \ .
}}
We will see below that the light contribution produces the expected Cardy behavior, whereas the contribution of the heavy operators we can estimate using the HKS bound. First, we estimate $\left| Z_H\left(4\pi^2 \over \beta + i t  \right) \right| \leq Z_H\left( 4\pi^2 \beta \over \beta^2 + t^2 \right)$ by removing phases. Then the RHS of the HKS bound \HKSb\ diverges exponentially as $t\to \infty$ when applied to $Z_H\left( 4\pi^2 \beta \over \beta^2 + t^2 \right)$. Therefore we require that $\widehat \phi_\pm(t)$ is decaying sufficiently rapidly at $t\to \infty$ so that the integrals in \boundsBBB\ converge.

One simple choice is to take $\widehat \phi_\pm(t)$ with support in a bounded region $t \in [- \Lambda_\pm, \Lambda_\pm]$. We then have
\eqn\estimateA{\eqalign{
\left| \int_{- \Lambda_-}^{\Lambda_-} d t ~ \widehat \phi_{-}(t) e^{-(\beta+i t) c/12} Z_H \left({4 \pi^2 \over \beta+ i t} \right) \right| &\leq e^{- \beta c/12} Z_H \left({4 \pi^2 \beta \over \beta^2+ \Lambda_-^2} \right) \int_{- \Lambda_-}^{\Lambda_-} d t~ | \widehat \phi_-(t) |  , \cr
\left| \int_{- \Lambda_+}^{\Lambda_+} d t~ \widehat \phi_{+}(t) e^{-(\beta+i t) c/12} Z_H \left({4 \pi^2 \over \beta+ i t} \right) \right| &\leq e^{- \beta c/12} Z_H \left({4 \pi^2 \beta \over \beta^2+ \Lambda_+^2} \right) \int_{- \Lambda_+}^{\Lambda_+} d t~ | \widehat \phi_+(t) | , \cr
}}
where it was absolutely crucial that the theory under consideration is unitary. The contribution of the heavy operators can be bounded using the HKS bound \HKSb\ or \HKSa. Also rewriting the first term in \boundsB\ back in $\Delta$-space we have
\eqn\boundsBB{\eqalign{
& e^{\beta (\Delta - \delta)} \left( \int_{0}^{\infty} d \Delta' ~ \rho_0(\Delta') e^{-\beta \Delta'}  \phi_{-}(\Delta')   -e^{- \beta c/12} Z_H \left({4 \pi^2 \beta \over \beta^2+ \Lambda_-^2} \right) \int_{- \Lambda_-}^{\Lambda_-} d t~ | \widehat \phi_-(t) |  \right) \cr
&\leq \int_{\Delta - \delta}^{\Delta + \delta} dF(\Delta') \leq \cr
&  e^{\beta (\Delta + \delta)} \left( \int_{0}^{ \infty} d \Delta' ~ \rho_0(\Delta') e^{-\beta \Delta'} \phi_{+}(\Delta')   + e^{- \beta c/12} Z_H \left({4 \pi^2 \beta \over \beta^2+ \Lambda_+^2} \right) \int_{- \Lambda_+}^{\Lambda_+} d t~ | \widehat \phi_+(t) |\right) \ .
}}

We do not know what is the best choice of $\phi_{\pm}(\Delta')$ within the class of functions with the Fourier transform of finite support and satisfying \functions\ that make the bounds optimal.  A simple and convenient choice is 
\eqn\choice{\eqalign{
\phi_+(\Delta') &=\left( {\sin {\Lambda_+ \delta \over 4} \over {\Lambda_+ \delta \over 4}  } \right)^{-4} \left( {\sin {\Lambda_+ (\Delta'-\Delta) \over 4} \over {\Lambda_+ (\Delta'-\Delta) \over 4}  } \right)^4 \ , \cr
\phi_-(\Delta') &=\left( {\sin {\Lambda_- (\Delta'-\Delta) \over 4} \over {\Lambda_- (\Delta'-\Delta) \over 4}  } \right)^4 \left(1 - \left({\Delta' - \Delta \over \delta} \right)^2 \right) \ .
}}
Note that these functions indeed satisfy \functions\ and their Fourier transform has a bounded support. Moreover, for this particular choice we have  $ \int_{- \Lambda_+}^{\Lambda_+} d t~ | \widehat \phi_+(t) | = 1$. Similarly, for ${1 \over \delta^2 \Lambda^2} < {1 \over 12}$ we have $ \int_{- \Lambda_-}^{\Lambda_-} d t~ | \widehat \phi_-(t) | = 1$. These are the values relevant for our finite $\Delta$ results in the 2d Ising section.

\subsec{Bounds at large $\Delta$}

The bound \boundsBB\ substantially simplifies in the limit $\Delta \gg 1$. Below we will see that in this case the optimal choice is $\beta = \pi \sqrt{c\over 3 \Delta} \ll 1$. Using the HKS bound we can show that the second terms in \boundsBB\ proportional to $Z_H$ are subleading for $\Lambda_{\pm} < 2 \pi$. Indeed we get 
\eqn\estimate{
 e^{\beta \Delta}Z_H \left({4 \pi^2 \beta \over \beta^2+ \Lambda_\pm^2} \right) \sim  e^{\beta \Delta} e^{{\pi^2 c \over 3 \beta}  ({ \Lambda_\pm \over 2 \pi})^2 } \sim \rho_{0}(\Delta)^{1+{1 \over 2} \left( ({ \Lambda_\pm \over 2 \pi})^2 - 1 \right)} ,
}
which will be subleading for $\Lambda_\pm< 2 \pi$ (we will see it momentarily below).  Therefore we get the bound at large $\Delta$
\eqn\boundC{\eqalign{
 &e^{\beta (\Delta - \delta)} \int_{0}^{\infty} d \Delta' \ \rho_0(\Delta')  \phi_-(\Delta') e^{- \beta \Delta'} \cr
 &\leq \int_{\Delta - \delta}^{\Delta + \delta} dF(\Delta') \leq \cr
 & e^{\beta (\Delta + \delta)} \int_{0}^{\infty} d \Delta' \ \rho_0(\Delta')  \phi_+(\Delta') e^{- \beta \Delta'} \ .
}}
The integrals can be computed by the saddle point approximation and give
\eqn\boundD{\eqalign{
&c_-  \rho_0(\Delta) \leq {1 \over 2 \delta} \int_{\Delta - \delta}^{\Delta + \delta} dF(\Delta') \leq c_+ \rho_0(\Delta) \ , \cr
&c_{\pm}  = {1 \over 2} \int_{- \infty}^{\infty} d x~ \phi_\pm(\Delta + \delta x) \ .
}}
We see that dropping the terms \estimate\ is indeed justified for $\Lambda_\pm <2\pi$. The explicit integration of \choice\ gives
\eqn\cpm{\eqalign{
c_+ = {\pi \over 3} {(\delta \Lambda_+/4)^3 \over [\sin (\delta \Lambda_+ /4) ]^4} \ , \qquad
c_- = {4\pi \over 3 (\delta \Lambda_-)^3} [(\delta \Lambda_-)^2 - 12]   \ .
}}
Note that for $\delta$ such that $c_- > 0$ we have to have at least one operator in the interval $[\Delta - \delta, \Delta+\delta]$ since in this case
\eqn\boundE{
\int_{\Delta - \delta}^{\Delta + \delta} dF(\Delta') > 0 .
}
This happens if
\eqn\cmin{
\delta^2 > {12 \over \Lambda_-^2} > {3\over \pi^2} \equiv \delta_{gap}^2 \ ,
}
where we also used the assumption $\Lambda_- < 2\pi$ to drop the term \estimate. That is for the simple choice of functions \choice\ we get $\delta_{gap}^2 = {3 \over \pi^2}$, which is to say that every modular invariant partition function has to have at least one operator within the window of size $2 \delta_{gap} = {2\sqrt{3} \over \pi} \approx 1.1$ at large $\Delta$. Of course, this is completely trivial in 2d CFTs due to the Virasoro descendants. However, in section 6 we will see that the same argument applies to Virasoro primaries as well provided $c>1$ and with the same result. It is natural to conjecture that the maximum allowed spacing between Virasoro primairy operators is in fact $1$.

Similarly, keeping $\delta $ arbitrary we can optimize over $0< \Lambda_{\pm} < 2\pi$ to get the tightest possible bound \boundD. For the lower bound the result is
\eqn\LboundF{\eqalign{
 {F(\Delta+\delta) - F(\Delta-\delta) \over 2\delta}  &\geq {4\pi \over 27}  \rho_0(\Delta) \approx 0.46  \rho_0(\Delta)  \ , \qquad \delta \geq {3\over \pi} \ ,  \cr
 {F(\Delta+\delta) - F(\Delta-\delta) \over 2\delta}  & \geq  {2(\delta^2 - \delta_{min}^2)\over 3\delta^3}   \rho_0(\Delta) \ , \qquad  \delta < {3\over \pi} \ .
}}
and for the upper bound
\eqn\UboundF{\eqalign{
{F(\Delta+\delta) - F(\Delta-\delta) \over 2\delta} &\leq {\pi \over 3 } {(a_*/4)^3 \over \sin(a_*/4)^4} \rho_0(\Delta) \approx 2.02  \rho_0(\Delta) \ , \qquad \delta \geq {a_*\over 2\pi} \ ,  \cr
 {F(\Delta+\delta) - F(\Delta-\delta) \over 2\delta} &\leq {\pi \over 3 } {(\pi \delta/2)^3 \over \sin(\pi \delta /2)^4}  \rho_0(\Delta) \ , \qquad \qquad \qquad \delta < {a_*\over 2\pi} \ ,
}}
where $a_*$ is the positive solution of the equation
\eqn\astar{
a_* = 3\tan (a_*/4) , \qquad a_* \approx 3.38 \ .
}
This bounds the number of ``outliers'' and shows what is the maximal local deviation of the density of operators from the Cardy distribution. Note that \LboundF, \UboundF\ already imply Cardy formula in the sense of entropies
\eqn\entropyCardy{
S_\delta(\Delta) = \log \int_{\Delta - \delta}^{\Delta + \delta} dF(\Delta') = 2\pi\sqrt{c\Delta \over 3} +  {1\over 4}\log \left( c\over 48  \Delta^3  \right)+ \log{2\delta} +  s(\delta, \Delta) \ , 
}
where $s$ is of $O(1)$ and can be bounded from \LboundF, \UboundF. We find
\eqn\sboundB{\eqalign{
\log {2(\delta^2 - \delta_{min}^2)\over 3\delta^3}     &\leq s(\delta, \Delta) \leq \log \left( {2 \over 3\delta } {(\pi \delta/2)^4 \over \sin(\pi \delta /2)^4}  \right)  \ , \qquad 0\leq \delta \leq {a_*\over 2\pi} \ , \cr 
\log {2(\delta^2 - \delta_{min}^2)\over 3\delta^3}     &\leq s(\delta, \Delta) \leq \log \left(   {\pi \over 3 } {(a_*/4)^3 \over \sin(a_*/4)^4} \right)   \ , \qquad {a_*\over 2\pi} \leq \delta \leq {3\over \pi} \ , \cr 
-0.76  \approx \log {4\pi \over 27}  &\leq s(\delta, \Delta) \leq \log\left(  {\pi \over 3 } {(a_*/4)^3 \over \sin(a_*/4)^4} \right) \approx 0.70   , \quad \delta \geq {3\over \pi} \ .
}}
The formula \entropyCardy\ is valid up to corrections suppressed at large $\Delta$. The $O(1)$ contribution $s(\delta,\Delta)$ is generically an oscillating function of $\Delta$. We will observe this explicitly in the 2d Ising model.  The bounds \sboundB\ are plotted in the \sbound.

It would be interesting to find the optimal bounds on the local density of operators by a better choice of $\phi_{\pm}$. To reiterate, in our argument these obey two defining properties: they satisfy \functions; they have a finite support in Fourier space \estimate.\foot{The latter can be slightly relaxed: it is sufficient to assume rapid decay at $t\to \infty$ so that the integrals in \boundsB\ converge after using the HKS bound. We have not explored this possibility.} Let us also emphasize that the bounds \boundsB, \boundsBB\ are applicable at finite $\Delta$ as well. In this case we should simply keep the terms \estimateA\ which we can estimate using the HKS bound.

\newsec{Proof of the Theorem}

In the previous section we investigated a local bound on the number of operators in a 2d CFT. In this section we derive a better bound for the case $\delta \gg 1$. In particular we show that if $\Delta \gg 1$ then averaging $\rho(\Delta)$ over operators in the region $[\Delta - \delta, \Delta+\delta]$ with $\delta \sim \Delta^\alpha$ for some $\alpha>0$ produces the fixed asymptotic identical to the one given by the crossing kernel $\rho_0(\Delta)$ with the controlled error \sboundpm. As mentioned in the introduction it follows from the theorem \introTrueCardy. We prove \introTrueCardy\ in this section which we repeat for convenience here
\eqn\theorem{
F(\Delta) \equiv \int_0^\Delta d\Delta' ~ \rho(\Delta') ={1\over 2\pi} \left( 3\over c\Delta \right)^{1/4} e^{2\pi \sqrt{{c\over 3} \Delta}} \left[ 1 + O(\Delta^{-1/2}) \right], \qquad \Delta \to \infty \ .
}

Few comments are in order. Note that by doing a naive inverse Laplace transform of the vacuum contribution, using the saddle point approximation, and integrating over $\Delta$ one would arrive at the correct estimate for $F(\Delta)$, namely \theorem.  Using the saddle point approximation to make a statement about $\rho(\Delta)$ itself however is not correct. It would be also incorrect to use the saddle point approximation to compute further corrections to $F(\Delta)$, beyond \theorem.

Let us introduce the difference between the Laplace transform of the physical density of states $\rho(\Delta)$ and the crossing kernel $\rho_0(\Delta)$
\eqn\diff{
\delta {\cal L}(\beta) = {\cal L}_\rho(\beta) - {\cal L}_{\rho_0} (\beta), \qquad 
\delta \rho(\Delta) = \rho(\Delta) - \rho_0(\Delta) \ .
}

\ifig\contour{Integration contour in the complex temperature $z$-plane. We integrate the modular invariance equation \modular\ along the vertical segment $C_+$ to derive the bound on the integrated spectral density.}{\epsfxsize2.7in\epsfbox{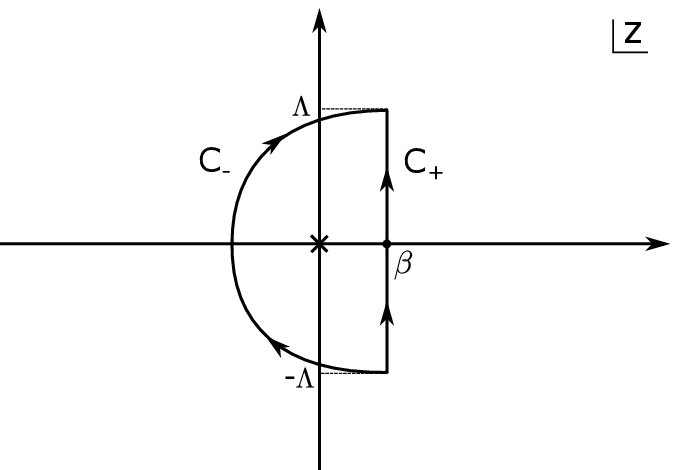}}

The main idea is to apply a linear functional to the modular invariance equation \modular\ that produces the theta-function $\theta(\Delta-\Delta')$ that we want plus terms which we can easily estimate. A convenient choice of the functional is
\eqn\linfunc{
{1\over 2\pi i }\int_{\beta - i \Lambda}^{\beta + i \Lambda} {dz \over z}  ~\left[ \Lambda^2 + (z - \beta)^2 \right] ~ e^{z \Delta } ~ \delta {\cal L}(z) \ ,
}
where the integration contour is the interval $C_+ = \{ {\rm Re}~ z = \beta, -\Lambda < {\rm Im}~z < \Lambda \}$ as indicated on the figure \contour. The parameters $\Lambda, \beta, \Delta$ are so far arbitrary in \linfunc. The polynomial in the numerator of \linfunc\ is chosen to be such that it vanishes at the ends of the interval $C$, which will be helpful in estimates below. 

On the one hand we can estimate \linfunc\ using modular invariance. Inserting the definition of the Laplace transform and swapping the order of integrations we have
\eqn\contdef{\eqalign{
&{1\over 2\pi i }\int_{\beta - i \Lambda}^{\beta + i \Lambda} {dz \over z}  ~{ \Lambda^2 + (z - \beta)^2  \over \Lambda^2 + \beta^2} ~ e^{z \Delta } ~ \delta {\cal L}(z) = 
   \int_0^\infty d\Delta'  ~ \delta \rho(\Delta') G(\Delta - \Delta'),  \cr
& G(\nu) = {1\over 2\pi i }\int_{\beta - i \Lambda}^{\beta + i \Lambda} {dz \over z} ~{ \Lambda^2 + (z - \beta)^2  \over \Lambda^2 + \beta^2} ~ e^{ -\nu z }  , \qquad \nu \equiv  \Delta' - \Delta
}}

Now the idea is to deform the contour $C_+$ in the last integral in \contdef\ either to the left or to the right for $\Delta' <\Delta$ or $\Delta' > \Delta$ respectively in order to make the exponential factor $e^{ (\Delta - \Delta')z}$ smaller. When we deform to the left we also pick up the residue at $z=0$. We have
\eqn\Gdefcont{
G(\nu) =\theta(-\nu) + \theta(\nu) G_+(\nu) + \theta(-\nu) G_-(\nu) ,
}
where $G_{\pm}(\nu)$ refer to the integrals over the arcs $C_{\pm}$, see \contour.

We can use \Gdefcont\ to rewrite the equation \contdef\ as follows 
\eqn\contdefA{\eqalign{
\int_0^\Delta d\Delta' ~ \delta \rho(\Delta') = &{1\over 2\pi i }\int_{\beta - i \Lambda}^{\beta + i \Lambda} {dz \over z}  ~{ \Lambda^2 + (z - \beta)^2  \over \Lambda^2 + \beta^2} ~ e^{z \Delta } ~ \delta {\cal L}(z)  \cr 
&- \int_0^\infty d\Delta' ~ \delta \rho(\Delta') [\theta(\Delta' - \Delta) G_+(\Delta - \Delta') + \theta(\Delta - \Delta') G_-(\Delta - \Delta')]
}}
In appendix A we show that\foot{The overall coefficient in this estimate is not optimal and can be improved, but it will be enough for our purposes.} 
\eqn\Gbound{
|G_\pm(\nu) | \leq 2 e^{-\beta \nu} \min[1, (\Lambda \nu)^{-2}] \ .
}
Therefore we can bound \contdefA\ as follows
\eqn\lemma{\eqalign{
\left| \int_0^\Delta d\Delta' ~ \delta \rho(\Delta')  \right| \leq &\left| {1\over 2\pi i }\int_{\beta - i \Lambda}^{\beta + i \Lambda} {dz \over z}  ~{ \Lambda^2 + (z - \beta)^2  \over \Lambda^2 + \beta^2} ~ e^{z \Delta } ~ \delta {\cal L}(z)  \right| \cr 
&+  2 e^{\beta \Delta}  \int_0^\infty dF(\Delta') ~e^{-\beta \Delta' } \min\left[1 , {1\over \Lambda^2 (\Delta' - \Delta)^2} \right] \ ,\cr
&+  2  e^{\beta \Delta} \int_0^\infty dF_0(\Delta') ~ e^{-\beta \Delta' } \min\left[1 , {1\over \Lambda^2 (\Delta' - \Delta)^2} \right] \ , 
}}
where we used the fact that $| \delta \rho(\Delta') | \leq \rho(\Delta') + \rho_0(\Delta')$ .

In the formula above $\beta$ is an arbitrary parameter. We would like to choose it to optimize the bound. We will show below that in order to prove \introTrueCardy\ the correct choice is to set
\eqn\betachoice{
\beta = \pi \sqrt{c\over 3\Delta}  \ .
} 
Let us emphasize that the bound \lemma\ is valid for finite $\Delta$. In particular we can use the HKS bound to estimate the first term in the RHS of \lemma\ and the local bound from the previous section to bound the second term. Below we investigate \lemma\ in the large $\Delta$ limit.

To estimate the third integral in the RHS of \lemma\ we use the asymptotic \naiveCardy
\eqn\integral{
e^{\beta \Delta} \int_0^\infty d\Delta' ~ \Delta'^{-3/4} e^{2\pi \sqrt{c\Delta' \over 3} - \beta \Delta'} 
\min[ 1,  (\Delta - \Delta')^{-2}]
= O \left(  \Delta^{-3/4} e^{2\pi \sqrt{c\Delta \over 3} } \right) \ .
}
The saddle here is at $\Delta' = \Delta$. Notice the importance of $\min[ 1,  (\Delta - \Delta')^{-2}]$. The second argument suppresses the integral over fluctuations $x = \Delta'-\Delta$ at large $x$ to produce the correct prefactor in the RHS of \integral. While the first argument cuts off the integral at small $x$ and makes it convergent there.

To estimate the second integral in the RHS of \lemma\ we split it into three parts $I_1,I_2, I_3$
\eqn\intsplit{
I_1+I_2+I_3 = \left( \int_0^{\Delta - \Delta^{3/8}} + \int_{\Delta - \Delta^{3/8}}^{\Delta + \Delta^{3/8}}+\int_{\Delta + \Delta^{3/8}}^\infty \right) d\Delta' ~\rho(\Delta') e^{\beta(\Delta - \Delta')} \min[ 1,(\Delta - \Delta')^{-2}] 
}
We would like to show that all three terms are of $O\left( \Delta^{-3/4} e^{2\pi \sqrt{c \Delta \over 3}} \right)$ separately. For $I_1$ we have
\eqn\ioneest{\eqalign{
I_1 &= \int_0^{\Delta - \Delta^{3/8}} d\Delta' ~\rho(\Delta') e^{\beta(\Delta - \Delta')} (\Delta - \Delta')^{-2}  \cr 
&\leq  \Delta^{-3/4} e^{\beta \Delta} \int_0^{\Delta - \Delta^{3/8}} d\Delta' ~\rho(\Delta') e^{- \beta \Delta'}  \cr 
&\leq  \Delta^{-3/4} e^{\beta \Delta} {\cal L}_\rho (\beta) = 
 O(\Delta^{-3/4} e^{2\pi \sqrt{c\Delta \over 3} }) \ ,
}}
where we used monotonicity of $(\Delta - \Delta')^{-2}$ in the first line and ${\cal L}_\rho(\beta) = O(e^{\pi^2 c \over 3\beta})$ and \betachoice\ in the third line. In particular, \ioneest\ shows that we chose to split the integral as in \intsplit\ in order to produce the correct prefactor in \ioneest\ $(\Delta - \Delta')^{-2} \Big|_{\Delta' = \Delta - \Delta^{3/8}} = \Delta^{-3/4}$. Similarly, $I_3$ is estimated to be of the same order
\eqn\ithreeest{\eqalign{
I_3 &= \int_{\Delta + \Delta^{3/8}}^\infty d\Delta' ~\rho(\Delta') e^{\beta(\Delta - \Delta')} (\Delta - \Delta')^{-2} \leq \Delta^{-3/4} e^{\beta \Delta}  {\cal L}_\rho(\beta)  = O(\Delta^{-3/4} e^{2\pi \sqrt{c\Delta \over 3} }) .
}}

Finally, we need to estimate $I_2$. We will do so using a local bound from the previous section
\eqn\localboundlarge{
F(\Delta +\delta) - F(\Delta - \delta) = O\left(\Delta^{-3/4} e^{2\pi \sqrt{c\Delta \over 3}}\right). 
}
We further split the integral $I_2$ into 
\eqn\itwoest{
I_2 = \left( \underbrace{ \int_{\Delta - \Delta^{3/8}}^{\Delta - 1}  }_{i_1}+ \underbrace{ \int_{\Delta - 1}^{\Delta + 1} }_{i_2}+ \underbrace{ \int_{\Delta + 1}^{\Delta + \Delta^{3/8}} }_{i_3}\right) d\Delta' ~\rho(\Delta') e^{\beta(\Delta - \Delta')} \min[ 1,(\Delta - \Delta')^{-2}]  
}
To estimate $i_{1,2,3}$ we split the integrals into small windows of $\Delta'$ in each of which we can apply \localboundlarge
\eqn\iiest{\eqalign{
i_1 &=  \int_{\Delta - \Delta^{3/8}}^{\Delta - 1}d\Delta' ~\rho(\Delta') e^{\beta(\Delta - \Delta')} (\Delta - \Delta')^{-2}  \cr 
&=\sum_{k=2}^{\Delta^{3/8}}  \int_{\Delta - k}^{\Delta - k+1} d\Delta' ~\rho(\Delta') e^{\beta(\Delta - \Delta')} (\Delta - \Delta')^{-2}  \cr 
& \leq \sum_{k=2}^{\Delta^{3/8}} { e^{\beta k} \over (k-1)^2} \int_{\Delta - k}^{\Delta - k+1} d\Delta' ~\rho(\Delta') =\sum_{k=2}^{\Delta^{3/8}} { e^{\beta k} \over (k-1)^2} [F(\Delta -k+1) - F(\Delta - k) ] \cr 
&= O\left(\Delta^{-3/4} e^{2\pi \sqrt{c\Delta \over 3}}  \sum_{k=2}^{\Delta^{3/8}} { e^{\beta k} \over (k-1)^2} \right) = O\left(\Delta^{-3/4} e^{2\pi \sqrt{c\Delta \over 3}}\right)
}}
where we used \localboundlarge\ and \betachoice. The integral $i_3$ is estimated in a similar fashion. Finally,
\eqn\iitwoest{
i_2 = \int_{\Delta - 1}^{\Delta + 1}d\Delta' ~\rho(\Delta') e^{\beta(\Delta - \Delta')} \leq e^\beta [F(\Delta +1) - F(\Delta-1)] = O\left(\Delta^{-3/4} e^{2\pi \sqrt{c\Delta \over 3}}\right)
}
This finishes the estimate of \intsplit.

The last step is to estimate the first term in the RHS of \lemma. To this we need to use the modularity condition\foot{Here we imagine subtracting a finite number of light operators in $\delta {\cal L}$ below $\Delta_H > c/12$. This does not affect previous estimates since such light operators would contribute terms analogous to the third term in the RHS of \lemma\ and would give exponentially small corrections to Cardy growth. }
\eqn\modularity{
\left| \delta {\cal L}(z) \right| = \left| e^{- z c \over 12} Z_{H} \left({4 \pi^2 \over z} \right) \right| \leq e^{- {\rm Re}[z]c/12} Z_{H}\left({4 \pi^2 {\rm Re}[z]\over |z|^2} \right) ,
}
which we can estimate using the vacuum contribution in the dual channel.
We get
\eqn\estimatefirstlemma{\eqalign{
&\left| {1\over 2\pi i }\int_{\beta - i \Lambda}^{\beta + i \Lambda} {dz \over z}  ~{ \Lambda^2 + (z - \beta)^2  \over \Lambda^2 + \beta^2} ~ e^{z \Delta } ~ \delta {\cal L}(z)  \right| \cr
& \leq {1 \over 2 \pi} \int_{- \Lambda}^{\Lambda} {d t \over | \beta + i t | } {\Lambda^2 + t^2 \over \Lambda^2 + \beta^2} e^{\beta (\Delta-{c \over 12}) } Z_{H}\left({4 \pi^2 \beta \over \beta^2 + t^2} \right)  \cr
&\leq {2 \pi \Lambda^3 e^{\beta (\Delta-{c \over 12}) } \over \beta (\Lambda^2 + \beta^2)} Z_H \left({4 \pi^2 \beta \over \beta^2 + \Lambda^2} \right) = O \left(\Delta^{1/2} e^{\pi \sqrt{c\Delta \over 3} \left(1 + ({\Lambda \over 2 \pi})^2 \right)} \right)
}}
where in the second line we used monotonicity of $Z_H$ and therefore assumed that $\Delta_H > {c \over 12}$. In the third line we estimated $Z_H$ using the vacuum contribution in the dual channel. Choosing $\Lambda < 2 \pi$ we see that this term is sub-leading. This finishes the proof of \theorem.

\newsec{Virasoro Primaries}

The analysis in previous sections can be readily generalized to the density of Virasoro primary operators. Let's consider $c>1$ so that there are infinitely many such operators. In this case Virasoro characters are simply related to the Dedekind function and the partition function takes the form, see e.g.  \HellermanBU,
\eqn\Virpartfunc{
Z(\beta) = |\eta (\tau)|^{-2} e^{\beta {c - 1 \over 12}} \left[ (1- e^{-\beta})^2 + \sum_{n=1}^\infty d_n^{\rm Vir} e^{-\beta \Delta_n} \right] \ ,
}
where $\tau = i\beta / 2\pi$, $d_n^{\rm Vir}$ is the degeneracy of a Virasoro primary $\Delta_n$ and the sum goes over all primaries except the vacuum $\Delta_n >0$. Let's define the density of Virasoro primaries
\eqn\virdensdef{\eqalign{
&\rho^{\rm Vir}(\Delta) = \sum_{n=1}^\infty  d_n^{\rm Vir} \delta (\Delta - \Delta_n) \ . 
}}
The crossing kernel for the vacuum is given by 
\eqn\virnaivedens{\eqalign{
\rho_0^{\rm Vir}(\Delta) &= f(\Delta, 0) - 2f(\Delta, 1) + f(\Delta, 2)  \ , \cr 
f(\Delta, x) &= 2\pi ~ 
I_0\left( 4\pi \sqrt{ \left({c-1 \over 12}-x \right) \left( \Delta - {c-1 \over 12} \right) } \right) 
 \theta\left(\Delta -{c-1\over 12} \right) - \delta(\Delta -x) \ ,
}}
so that it reproduces the vacuum contribution in the dual channel
\eqn\CrossKerLap{\eqalign{
&|\eta (\tau)|^{-2} e^{\beta {c - 1 \over 12}} \left[ (1- e^{-\beta})^2 + {\cal L}_{\rho_0^{\rm Vir}}(\beta) \right] = 
|\eta (\tau')|^{-2} e^{\beta' {c - 1 \over 12}}  (1- e^{-\beta'})^2 \ , \cr 
& \beta'  \equiv {4\pi^2 \over \beta}, \qquad \tau' \equiv - {1\over \tau} , \qquad {\cal L}_\rho(\beta) = \int_0^\infty d\Delta ~ \rho(\Delta) e^{-\beta \Delta}  \ .
}}

\subsec{Local bounds on the number of Virasoro primaries}

\noindent We can derive bounds analogous to \boundsB, \boundsBB. Essentially the same argument gives\foot{Here, as in \boundsB, it is implied that the crossing kernel $\rho_0^{\rm Vir}$ is for all light operators entering $Z_L$. But again in the large $\Delta$ analysis below the vacuum contribution will be dominant.}  
\eqn\boundsVir{\eqalign{
& e^{\beta (\Delta - \delta)} \int_{0}^{\infty} d \Delta' ~ \rho_0^{\rm Vir}(\Delta') e^{-\beta \Delta'}  \phi_{-}(\Delta')   \cr 
&- e^{\beta \left(\Delta - \delta - {c-1 \over 12} \right) } Z_H\left({4 \pi^2 \beta \over \beta^2+ \Lambda_-^2} \right) \int_{- \Lambda_-}^{ \Lambda_-} d t ~ \hat \phi_{-}(t)  \left| \eta\left( i(\beta+ it) \over 2\pi \right) \right|^2   \cr
&\leq \int_{\Delta - \delta}^{\Delta + \delta} dF(\Delta') \leq \cr
& e^{\beta (\Delta + \delta)}\int_{0}^{\infty} d \Delta' ~ \rho_0^{\rm Vir}(\Delta') e^{-\beta \Delta'}  \phi_{+}(\Delta')    \cr 
&+ e^{\beta \left(\Delta + \delta - {c-1 \over 12} \right) } Z_H\left({4 \pi^2 \beta \over \beta^2+ \Lambda_+^2} \right) \int_{- \Lambda_+}^{ \Lambda_+} d t ~ \hat \phi_{+}(t)  \left| \eta\left( i(\beta+ it) \over 2\pi \right) \right|^2   \ .
}}
The HKS bound for Virasoro primaries can also be derived and takes the form
\eqn\VirHKS{\eqalign{
&Z_H \leq   {\beta \over 2\pi} e^{-(\beta - \beta') (\Delta_H - {c-1 \over 12}) }{ Z_L - Z_L'   \over 1 - {\beta \over 2\pi} e^{-(\beta - \beta') (\Delta_H - {c-1\over 12} ) }} , \qquad \beta \geq 2\pi \ , \cr 
&Z_H \leq  { Z_L' - Z_L  \over 1 - {\beta' \over 2\pi}e^{-(\beta' - \beta) (\Delta_H - {c-1\over 12})}} , \qquad \beta \leq 2\pi \ .
}}
where $\Delta_H > {c-1 \over 12}$ and we split the partition function into light and heavy contributions
\eqn\Virsplit{\eqalign{
&Z_L =  |\eta (\tau)|^{-2} e^{\beta {c - 1 \over 12}} \left[ (1- e^{-\beta})^2 + \sum_{0<\Delta_n <\Delta_H } d_n^{\rm Vir} e^{-\beta \Delta_n} \right] \ , \cr 
&Z_H=  |\eta (\tau)|^{-2} e^{\beta {c - 1 \over 12}} \sum_{\Delta_n \geq \Delta_H } d_n^{\rm Vir} e^{-\beta \Delta_n} \ .
}}
The large $\Delta$ analysis is identical to the section 4 and with essentially the same results. Namely we get
\eqn\VirlargeDelta{
\int_{\Delta - \delta}^{\Delta + \delta} dF(\Delta') >0 ,\qquad 2\delta > 2\delta_{gap} = 2 \sqrt{3\over \pi^2} 
}
with the choice \choice. That is the gap between Virasoro primaries at large scaling dimensions must be no larger than $2\sqrt{3\over \pi^2} \approx 1.1$. 

Repeating the rest of the argument from the section 4 we obtain the asymptotic of the microcanonical entropy for energy shells $\delta =O(1)$
\eqn\virMicroCardyA{\eqalign{
S_\delta^{\rm Vir}(\Delta) &\equiv \log \int_{\Delta - \delta}^{\Delta + \delta} d\Delta' ~\rho^{\rm Vir}(\Delta') \cr 
&= 2\pi \sqrt{ {c-1 \over 3 } \Delta} - {1\over 4} \log \left( {c-1 \over 3} \Delta \right) + \log(2\delta) + s^{\rm Vir}(\delta , \Delta) , \qquad \Delta \to \infty \ ,
}}
where $s^{\rm Vir}(\delta, \Delta)$ is again bounded as in \sbound.

\subsec{Cardy formula for Virasoro primaries}

\noindent The modular invariance dictates the behavior at high temperatures
\eqn\virhighT{
\int_0^\infty d\Delta ~ \rho^{\rm Vir}(\Delta) e^{-\beta \left( \Delta - {c-1 \over 12} \right)} =  {2\pi \over \beta} e^{{4\pi^2 \over \beta} {c-1 \over 12} } \left[ 1 + O\left( \max \left[ e^{-{4\pi^2 \over \beta}}, e^{-{4\pi^2 \over \beta} \Delta_1} \right] \right) \right], \qquad \beta \to 0 \ .
}
Then the tauberian theorem similar to \introTrueCardy\ takes the form
\eqn\virglobalCardy{
\int_0^\Delta d\Delta' ~\rho^{\rm Vir}(\Delta') = {1\over \pi} \left( 3 \over c-1 \right)^{3/4} \Delta^{1/4} e^{ 2\pi \sqrt{ {c-1 \over 3} \Delta } } \left[ 1 + O(\Delta^{-1/2} \right] \ .
}
Its proof is completely analogous to the proof of \introTrueCardy\ given in the section 5. From here we derive that the microcanonical entropy has the asymptotic \virMicroCardyA\ with $s(\delta,\Delta)$ given by
\eqn\virMicroCardy{\eqalign{
&s^{\rm Vir}(\delta, \Delta) = \log\left(  { \sinh  \pi \sqrt{ c-1 \over 3} {\delta \over \sqrt{\Delta} } \over  \pi \sqrt{ c-1 \over 3} {\delta \over \sqrt{\Delta} } } \right) + O(\Delta^{-\alpha}) , \qquad \delta \sim \Delta^\alpha, \quad \Delta \to \infty 
}}
for any $0 < \alpha \leq 1/2$.

\newsec{Holographic CFTs}

In this section we consider holographic 2d CFTs with a sparse spectrum \HartmanOAA\ in the limit $\Delta \sim c \to \infty$. The HKS sparseness condition \HartmanOAA\ states that $Z_L(\beta)$ is dominated by the vacuum state for $\beta > 2\pi$ and $c \to \infty$ in the sense that
\eqn\HKSsparse{
 \sum_{\Delta \leq \Delta_H} e^{-\beta \Delta}  = O(1), \qquad  \beta > 2\pi, \quad c \to \infty \ .
}
We again start with \boundsBB\ and consider the limit
\eqn\notlargec{
\Delta = c \left( {1 \over 12} + \e \right) \ , \qquad c \to \infty, \quad \e {\rm - fixed} \ .
}
In this limit the asymptotic of the vacuum crossing kernel is
\eqn\crosskerlargeC{
\rho_{0}(\Delta) = 
   {1\over 2 \cdot 3^{1/4}} c^{-1/2} \e^{-3/4} e^{2\pi c \sqrt{\e /3}}  \theta(\e)
 + \dots \ .
}
To optimize the first term in \boundsBB\ we choose
\eqn\betachoiceLargeC{
\beta = {\pi \over \sqrt{3 \e}} \ .
} 
As before we find that the second $Z_H$ term in \boundsBB\ is suppressed if $\Lambda_{\pm} < 2 \pi \sqrt{1 - {1 \over 12 \eps}}$. Therefore the bound \boundsBB\ can be dominated by the first term only for $\eps>{1 \over 12}$, i.e. for states with $\Delta > {c \over 6}$. In this case we drop the $Z_H$ terms, compute the first term in \boundsBB\ by the saddle approximation and get
\eqn\constraintlargeC{
e^{-{\pi \delta  \over \sqrt{3 \e}}} \rho_0(\Delta) c_- \leq  {1 \over 2 \delta} \int_{\Delta - \delta}^{\Delta + \delta} dF(\Delta') \leq e^{{\pi \delta  \over \sqrt{3 \e}}} \rho_0(\Delta) \t c_+ \ , \qquad c \to \infty, \quad  \e {\rm - fixed} 
}
In \constraintlargeC\ we tacitly assumed that the first term in \boundsBB\ is dominated by the vacuum. For the RHS of \constraintlargeC\ this relies on sparseness condition and we give more detail in appendix C. In particular, this means that we cannot compute the precise value of $\t c_+$ because it depends on the bound \HKSsparse. On the other hand In the LHS of \constraintlargeC\ we can simply drop operators above the vacuum since they give positive contribution.

The conclusion is that we have the asymptotic of the microcanonical entropy of states with energy of $O(c)$
\eqn\LargeCmicro{
\log \int_{\Delta - \delta}^{\Delta + \delta} d\Delta' \rho(\Delta') = 2\pi  \sqrt{{c \over 3}\left( \Delta - {c\over 12} \right)} - {1\over 2} \log c + O(1), \qquad \Delta > {c\over 6}, \quad c \to \infty 
}
for fixed $\delta > \delta_{gap} =  {\sqrt{3} \over \pi}$.

We can also consider large widths $\delta \sim c^\alpha, 0<\alpha < 1$. We estimate by splitting into intervals of $O(1)$
\eqn\LargeCdelta{\eqalign{
&\int_{\Delta - \delta}^{\Delta + \delta} d\Delta' ~ \rho(\Delta') = \sum_{k=1}^{2\delta - 1} \int_{\Delta - \delta + k -1}^{\Delta - \delta + k +1} d\Delta' ~\rho(\Delta')
}}
and applying the bound \constraintlargeC\ to each term. Both the upper and lower bounds are dominated by the largest exponent $k = 2\delta -1$ and are estimated to be
\eqn\LargeCLargedelta{
O\left( \sum_{k=1}^{2\delta - 1} c^{-1/2} e^{2\pi c \sqrt{{1 \over 3} \left( \e + {  k-\delta \over c} \right) }  }   \right)  
 =O\left(  c^{-1/2} e^{2\pi c \sqrt{{1 \over 3} \left( \e + {  \delta \over c} \right) }  }   \right) \ .
}
Therefore we have for the microcanonical entropy
\eqn\LargeCentropy{
S_{\delta}(\Delta) = \log \int_{\Delta - \delta}^{\Delta + \delta} d\Delta' \rho(\Delta') = 2\pi  \sqrt{ {c \over 3} \left( \Delta + \delta - {c\over 12} \right)}  - {1\over 2} \log c + O(1), \qquad c \to \infty \ ,
}
where $\delta \sim c^\alpha, 0 < \alpha<1$. For $0 < \alpha \leq {1\over 2}$ only the first term in the expansion of the square root dominates the error. For $1/2 < \alpha < 1$ more terms in the expansion of the square root give a contribution. Essentially, the formula \LargeCentropy\ states that the entropy is dominated by the states in an $O(1)$ window near the upper limit. In \HartmanOAA\ it was derived that $S_{\delta} (\Delta) = 2\pi c \sqrt{\e \over 3} + O(c^{\alpha}), 1/2 < \alpha < 1$. The formula \LargeCentropy\ extends their result to all $0 < \alpha < 1$ and computes corrections to it.

It would be interesting to reproduce our result for the microcanonical entropy \LargeCentropy\ from the direct bulk computation. The leading contribution to the on-shell action is insensitive to the ensemble choice, however the state of the quantum fields in the black hole background changes which should be taken into account when computing the corrections to the leading Cardy formula, see e.g. \refs{\BrownBQ,\MarolfLDL}.

Note that the logarithmic correction to the microcanonical entropy \LargeCentropy\ is completely universal. This feature of AdS black holes in AdS was observed in \SenDW\ (see section 5 in that paper) and is due to the fact that there are no translational zero modes in AdS. The situation is drastically different from flat space, where logarithmic corrections to the black hole entropy are sensitive to the low energy spectrum of the theory.

\newsec{Accessing Subleading Operators}

\noindent One way to access the subleading operators in the dual channel is the following. Consider the modular condition written as
\eqn\modularB{\eqalign{
&\int_0^{\Delta} d\Delta' ~ \delta \rho(\Delta') e^{-\beta (\Delta' - c/12)}  + \int_{\Delta}^\infty d\Delta' ~ \delta \rho(\Delta') e^{-\beta (\Delta' - c/12)} = Z \left({4 \pi^2 \over \beta}  \right) - e^{\pi^2 c \over 3 \beta} \ , \cr 
&\delta \rho(\Delta) \equiv \rho(\Delta) - \rho_0(\Delta) \ ,
}}
where we split the partition function into the contribution of light $\Delta' \leq \Delta$ and heavy $\Delta'>\Delta$ operators. Intuitively, it is clear that the partition is dominated by light (heavy) operators at small (high) temperatures. More precisely, we claim that at small temperatures modular invariance \modularB\ can be written as
\eqn\claimA{\eqalign{
&\int_0^{\Delta} d\Delta' ~ \delta \rho(\Delta') e^{-\beta (\Delta'-c/12)}    \cr 
&=Z \left({4 \pi^2 \over \beta}\right) - e^{\pi^2 c \over 3 \beta} +O\left(\Delta^{-3/4} e^{2\pi \sqrt{{c\over 3} \Delta} - \beta \Delta} \right), \quad \beta \geq \pi \sqrt{c\over 3 \Delta} \ , 
}}
while at high temperatures
\eqn\claimB{\eqalign{
&\int_{\Delta}^\infty d\Delta' ~ \delta \rho(\Delta') e^{-\beta (\Delta'-c/12)}    \cr 
&=Z \left({4 \pi^2 \over \beta}\right) - e^{\pi^2 c \over 3 \beta} +O\left(\Delta^{-3/4} e^{2\pi \sqrt{{c\over 3} \Delta} - \beta \Delta} \right), \quad \beta \leq \pi \sqrt{c\over 3 \Delta} \ . 
}}
Equivalently, at small (high) temperatures heavy (light) operators are suppressed
\eqn\claimC{\eqalign{
&\int_{\Delta}^\infty d\Delta' ~ \delta \rho(\Delta') e^{-\beta (\Delta'-c/12)}  = O\left(\Delta^{-3/4} e^{2\pi \sqrt{{c\over 3} \Delta} - \beta \Delta} \right), \quad \beta \geq \pi \sqrt{c\over 3 \Delta} \ , \cr 
&\int_0^{\Delta} d\Delta' ~ \delta \rho(\Delta') e^{-\beta (\Delta'-c/12)}  = O\left(\Delta^{-3/4} e^{2\pi \sqrt{{c\over 3} \Delta} - \beta \Delta} \right), \quad \beta \leq \pi \sqrt{c\over 3 \Delta} \ .
}}
Formulae \claimA\ - \claimC\ hold in the limit $\Delta \to \infty$. We derive them below. But first a few comments are in order. As we take $\beta \to \infty$ in \claimA\ the LHS is dominated by a few light operators, while the RHS, i.e. the dual channel, receives contribution from a large number of heavy operators entering $Z(4\pi^2/\beta)$. The error term is exponentially small in this case. Similarly in \claimB\ as we take $\beta \to 0$ an infinite number of heavy operators dominate the LHS, while a small number of light operators dominate in the RHS. Both cases are therefore consistent with the intuition that a light operator in one channel is reproduced by a large number of heavy operators in the dual channel. The most interesting case is the intermediate regime $\beta \sim \Delta^{-1/2}$ when both channels are dominated by light operators in the following sense. In this case we can tune $\beta$ so that a finite number of light operators beyond the vacuum contribute in the RHS of \claimA. Their effect is then reflected in the density of ``light" states $\Delta' < \Delta$ in the LHS of \claimA. This can be thought of as ``non-perturbative corrections'' to Cardy formula from operators beyond the vacuum and is discussed in more detail below.

Now let us derive \claimA\ - \claimC. Consider for example $\beta \geq \pi \sqrt{c\over 3 \Delta}  $ and the first estimate in \claimC. We write $\delta \rho(\Delta) = \p_{\Delta} \delta F(\Delta)$ and integrate by parts to get
\eqn\modularD{\eqalign{
\int_{\Delta}^\infty d\Delta' ~ \delta \rho(\Delta') e^{-\beta \Delta'} &= - \delta F(\Delta)e^{-\beta \Delta} +  \beta \int_{\Delta}^\infty d\Delta' ~ \delta F(\Delta')  e^{-\beta \Delta'}  \ , \cr  
\delta F(\Delta) & \equiv F_\rho(\Delta) - F_{\rho_0}(\Delta)   \ .
}}
We can estimate this using the error term in the Cardy formula \introTrueCardy. We get
\eqn\estimate{\eqalign{
- e^{- \beta \Delta}  \delta F(\Delta) &= O\left(\Delta^{-3/4} e^{2\pi \sqrt{{c\over 3} \Delta} - \beta \Delta} \right) , \cr
 \beta \int_{\Delta}^\infty d\Delta' ~ \delta F(\Delta') e^{-\beta \Delta'}  &= O\left( \beta \int_{\Delta}^\infty d\Delta' ~  \Delta'^{-3/4} e^{2\pi \sqrt{{c\over 3} \Delta'} }   e^{-\beta \Delta'}  \right) \ .
 }}
For $\beta > \pi \sqrt{c \over 3\Delta }$ the saddle point in the last integral is outside of the integration range and therefore it is dominated close to the lower limit $\Delta \to \infty$. As a result we get the first estimate in \claimC.

Similarly, the second estimate in \claimC\ is obtained by integration by parts and using Cardy formula \introTrueCardy.

The formulae \claimA, \claimB\ allow us to probe subleading operators in the dual channel. In particular, one might hope to test \claimA\ numerically for finite $\Delta$. We will do so in the 2d Ising model in the next section. Let's see what operators give contributions larger than the error term. Consider an operator with dimension $\Delta^*$ in the RHS of \claimA. Its contribution to the partition function in the dual channel takes the form $e^{{\pi^2 c \over 3 \beta} - {4 \pi^2 \over \beta} \Delta^*}$. The condition that it is greater than the error term is
\eqn\condition{
\Delta^*  \leq {c \over 12} \left( {\beta \over \pi \sqrt{c  \over 3 \Delta}} - 1 \right)^2 \ .
}
In particular, \condition\ implies that we have to scale $\beta \sim \Delta^{-1/2}$ if we would like to access a finite number of operators in the dual channel in the limit $\Delta \to \infty$.

To summarize, the partition function \claimA\ with the UV cut-off $\Delta$ and temperature $\beta > \pi \sqrt{c \over 3 \Delta }$ allows to systematically probe the operators in the dual channel satisfying \condition. We will test \claimA\ numerically in the 2d Ising model in section 6.

Finally, the formulae similar to \claimC\ for Virasoro primaries take the form
\eqn\virsubresult{\eqalign{
&\int_{\Delta}^\infty d\Delta ~ \delta \rho(\Delta) e^{-\beta \Delta}  = O\left(\Delta^{-1/4} e^{2\pi \sqrt{{c-1\over 3} \Delta} - \beta \Delta} \right), \quad \beta \geq \pi \sqrt{c-1\over 3 \Delta} \ , \cr 
&\int_0^{\Delta} d\Delta ~ \delta \rho(\Delta) e^{-\beta\Delta}  = O\left(\Delta^{-1/4} e^{2\pi \sqrt{{c-1\over 3} \Delta} - \beta \Delta} \right), \quad \beta \leq \pi \sqrt{c-1\over 3 \Delta} \ .
}} 
where $\delta \rho(\Delta) =\rho(\Delta)-\rho_0(\Delta) $.

\newsec{Example: 2d Ising}
\noindent In this section we check our results in the 2d Ising model. In particular, we will see that the error estimates are optimal. The partition function is given by \DiFrancescoNK

\eqn\Isingpart{
Z(\beta)= {1\over 2} \left| \theta_2(\tau) \over \eta(\tau) \right| +{1\over 2} \left| \theta_3(\tau) \over \eta(\tau) \right| +{1\over 2} \left| \theta_4(\tau) \over \eta(\tau) \right|  \ ,
}
where 
\eqn\thetadef{\eqalign{
&\eta(\tau) = q^{1/24} \prod_{n=1}^\infty (1-q^n), \qquad
\theta_2(\tau) = 2 q^{1/8} \prod_{n=1}^\infty (1-q^n)(1+q^n)^2, \cr 
&\theta_3(\tau) =  \prod_{n=1}^\infty (1-q^n)(1+q^{n-1/2})^2, \qquad 
\theta_4(\tau) =  \prod_{n=1}^\infty (1-q^n)(1-q^{n-1/2})^2 
}}
and we restrict to zero angular potential $q = e^{-\beta}$ as before and the central charge is $c= {1\over 2}$. Expanding the partition function in $q$ we can find degeneracies of operators.

\subsec{Unit operator}

\ifig\leadorder{$F_\rho(\Delta)$ (blue) and its smooth approximation $F_{\rho_0}(\Delta)$ (orange). To check the error term in \introTrueCardy\ we plot $(F_\rho(\Delta) - F_{\rho_0} (\Delta) ) \Delta^{3/4} e^{-2\pi \sqrt{{c\over 3} \Delta}}$ (inside the box). It is oscillating with a constant amplitude, as predicted by \introTrueCardy.}{\epsfxsize3.8in\epsfbox{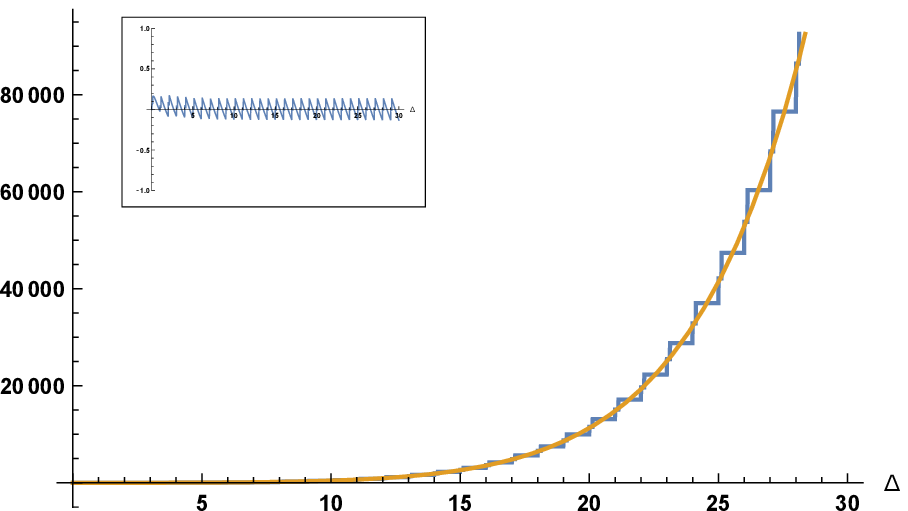}}

On \leadorder\ we plot the leading order and the error term for the moment $F_{\rho}(\Delta) = \int_0^\Delta d\Delta' ~ \rho(\Delta')$ and find perfect agreement with \theorem. In particular, it is clear from \leadorder\ that the error estimate is, in fact, optimal.

\subsec{$\sigma$ operator}

\noindent Now let's see how the effect of the first operator above the vacuum $\Delta_\sigma = {1\over 8}$ can be seen from the formula \claimA. According to \condition\ for $\sigma$ to give a contribution bigger than the error term and for $\Delta_\e = 1$ be smaller than the error term we require
\eqn\sigmaeps{
\Delta_\sigma \leq {c\over 12} \left( {\beta \over \pi \sqrt{c\over 3 \Delta}} -1 \right)^2 < \Delta_\e \ .
}
Inserting the numerical values we find that $\beta$ must be chosen in a window
\eqn\bwindow{
  {2.7 \pi \over \sqrt{6 \Delta}}  \approx {(1 + \sqrt{3}) \pi \over \sqrt{6 \Delta}} \leq \beta < {(1 + 2\sqrt{6}) \pi \over \sqrt{6 \Delta}} \approx  { 5.9\pi \over \sqrt{6 \Delta}} \ .
}
We take $\beta={4\pi \over \sqrt{6 \Delta}}$. The formula \claimA\ becomes
\eqn\bfour{\eqalign{
&\int^\Delta_0 d\Delta' ~e^{- \beta  \Delta' }\rho(\Delta') = 
 \int^\Delta_0 d\Delta' ~e^{ -\beta \Delta' }\rho_0(\Delta') +  \cr 
& + e^{- \beta c/ 12} e^{{\pi^2 c \over 3\beta} - {4\pi^2 \over \beta} \Delta_\sigma } 
  +
O \left(  \Delta^{-3/4} e^{2\pi \sqrt{{c \over 3} \Delta}  - \beta \Delta} \right) \ .
}}
Below we plot the contribution of $\sigma$-operator to \bfour\ and find perfect agreement. One can also plot the error term similarly to \leadorder.

\ifig\suberror{The difference $\int_0^\Delta d\Delta' e^{-\beta \Delta'} \delta \rho(\Delta')$ (blue) and $e^{- \beta c/ 12} e^{{\pi^2 c \over 3\beta} - {4\pi^2 \over \beta} \Delta_\sigma }$ (orange).}{\epsfxsize2.7in\epsfbox{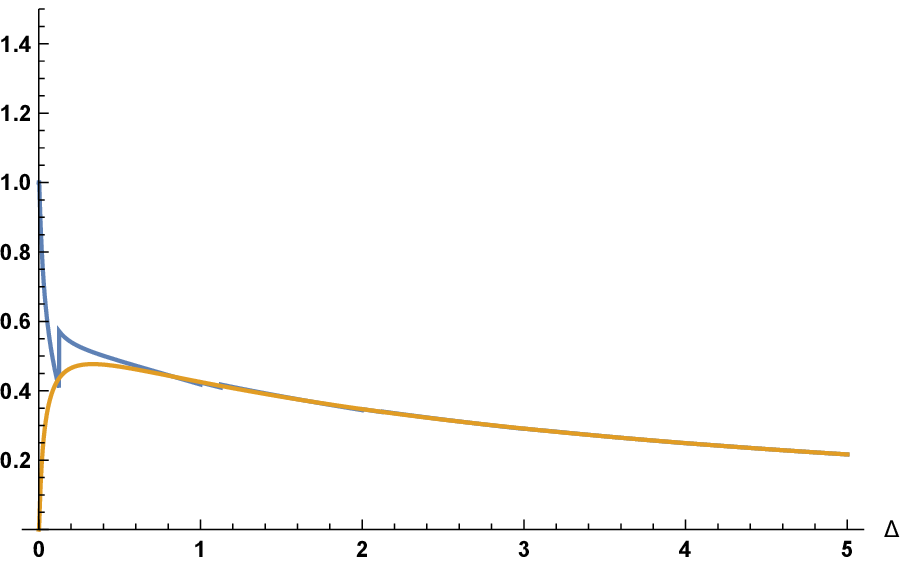}}

\subsec{Microcanonical Entropy}

As discussed in the main text the microcanonical entropy $S_{\delta}(\Delta)$ takes the universal form \IntroMicroCardy\ at high energies $\Delta \gg 1$.

\ifig\entropyIs{We plot $s(\delta,\Delta) $ (green) as defined in \IntroMicroCardy\ in the 2d Ising model for $\delta = 1.7$ as a function of $\Delta$. The straight lines correspond to the upper $s_+(\delta)$ (orange) and lower $s_-(\delta)$ (blue) asymptotic bounds. Dots represent the rigorous finite $\Delta$ bounds \boundsBB.}{\epsfxsize3.in\epsfbox{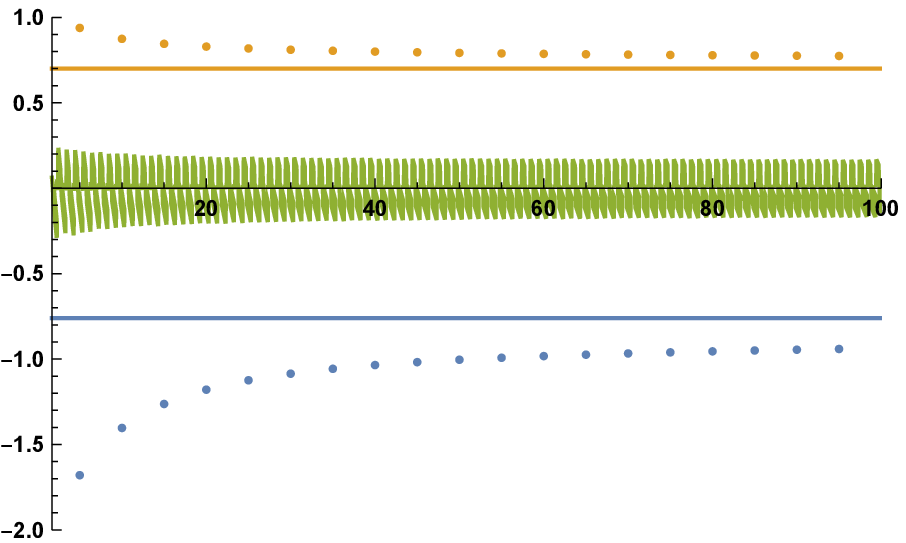}}

Here we explicitly plot the $O(1)$ correction $s(\delta,\Delta)$ to the leading behavior of the entropy in the 2d Ising model, see \entropyIs. In agreement with the general discussion we find that $s(\delta,\Delta)$ is an oscillating function with oscillations satisfying general bounds 
\sboundpm. Note that, strictly speaking, the bound \sboundpm\ was derived in the large $\Delta$ limit and here we plot it at finite $\Delta$. We present the finite $\Delta$ version of the bound \boundsBB\ on the \entropyIs\ as well. Since for 2d Ising model vacuum is the only operator with $\Delta < {c \over 12}$ we use only the vacuum contribution in $Z_L$ that enters the HKS bound. We then use the HKS bound to estimate $Z_H$.

\newsec{Example: Monster CFT}

Let us apply our bounds for the microcanonical entropy to monster CFT \refs{\FrenkelXZ,\WittenKT}. Recall that it describes a chiral CFT with $c=24$  
and the partition function that takes the form
\eqn\partitionfunctionMonster{
Z(q) = J(q) = {1 \over q} + 196884 q + 21493760 q^2 + ... \ .
}
In principle, nothing prevents us from deriving \virMicroCardy\ for chiral CFTs. We do not do this here. Instead, at zero angular potential and without imposing invariance under $\tau \to \tau +1$ we can interpret \partitionfunctionMonster\ as a partition function of a non-chiral CFT with $c=12$ that satisfies \modular. Therefore we can apply the asymptotic \IntroMicroCardyVir\ to it directly.

\ifig\entropyMfin{We plot $s^{{\rm Vir}}(\delta,\Delta) $ (green) versus $\Delta$ as defined in \IntroMicroCardyVir\ for the non-chiral Monster CFT partition function for $\delta = 2.4$ as a function of $\Delta$. The straight lines correspond to the upper $s^{{\rm Vir}}_+(\delta)$ (orange) and lower $s^{{\rm Vir}}_-(\delta)$ (blue) asymptotic bounds. We see that the actual microscopic entropy $s^{{\rm Vir}}(\delta,\Delta)$ is oscillatory, however, the amplitude of oscillations lays well within the asymptotic bounds for $\Delta$'s greater $\simeq 20$. Of course, as in the 2d Ising model the actual $\Delta$ bounds are weaker and are given by \boundsVir.}{\epsfxsize2.7in\epsfbox{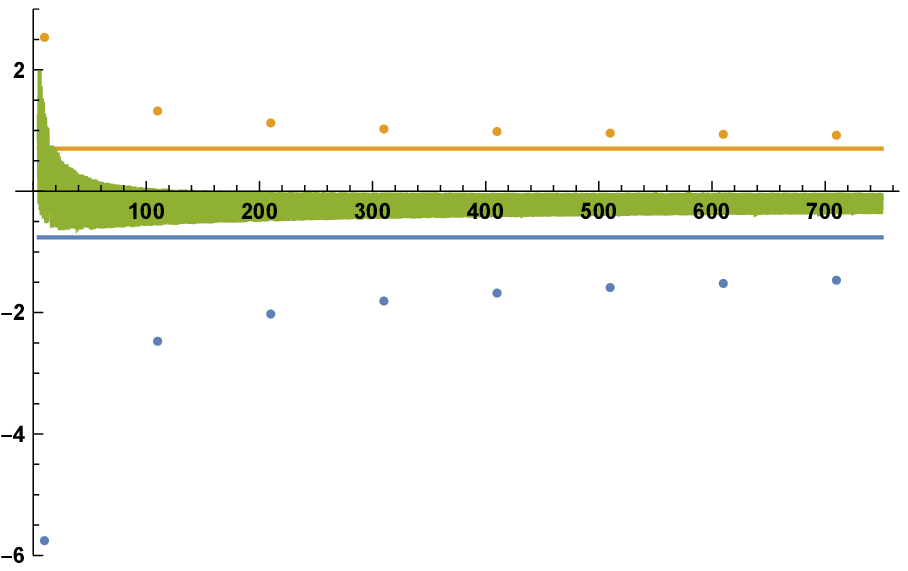}}

\ifig\entropyMMfin{We plot $s^{{\rm Vir}}(\delta,\Delta) $ (blue) as defined in \IntroMicroCardyVir\ for the non-chiral Monster CFT partition function for $\delta = \Delta^{0.4}$ as a function of $\Delta$. The orange line is given by the predicted universal correction $ \log { \sinh  \pi \sqrt{ c-1 \over 3} {\Delta^{0.4} \over \sqrt{\Delta} } \over  \pi \sqrt{ c-1 \over 3} {\Delta^{0.4} \over \sqrt{\Delta} } }$. In the box we plot $\left(s^{{\rm Vir}}(\Delta^{0.4},\Delta) -  \log { \sinh  \pi \sqrt{ c-1 \over 3} {\Delta^{0.4} \over \sqrt{\Delta} } \over  \pi \sqrt{ c-1 \over 3} {\Delta^{0.4} \over \sqrt{\Delta} } } \right) \Delta^{0.4}$ to check that the non-universal difference between the two curves is consistent with \IntroMicroCardyVir.
}
{\epsfxsize2.7in\epsfbox{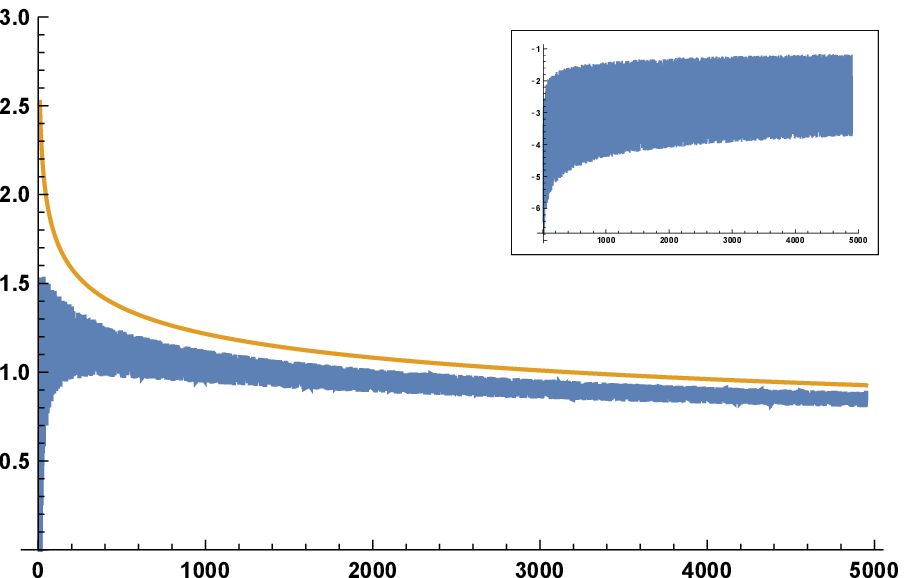}}

On \entropyMfin\ we see that for finite $\delta$ the difference between the actual microcanonical entropy and the large $\Delta$ expansion satisfies the expected bounds. We can also probe a subleading universal correction by taking $\delta = \Delta^\eps$. The result is presented on \entropyMMfin.

\newsec{Discussion}

In this paper we studied modular invariance of unitary $2d$ partition functions. Its most famous consequence is the Cardy formula \Cardy, which states that the density of states of unitary CFTs at high energies takes a simple, universal form. Needless to say there is a vast amount of work where the modular invariance of $2d$ partition functions is explored and variations of the Cardy formula are discussed. However, we have not found a rigorous derivation of the microcanonical entropy $S_{\delta}(\Delta)$ \microcan\ from the canonical entropy \CardyZ\ in the literature. We closed this gap by considering a set of linear functionals applied to the modularity condition that naturally appear in tauberian theory \Korevaar. The corresponding tauberian theorems are very general and not bounded to the discussion of partition functions in $2d$ CFTs. In particular, they are applicable to the higher-dimensional discussions of modular invariance \BelinYLL, warped $2d$ CFTs \refs{\DetournayPC,\SongTXA}, as well as to the thermal two-point function \HikidaKHG\ and the vacuum four-point functions \refs{\DasVEJ,\CollierEXN,\MaldacenaIUA\DasCNV\KusukiNMS-\KusukiWPA} (all of which can be studied using modular bootstrap tools in $2d$). It would be especially interesting to see if the methods used in this paper could shed light on the Eigenstate Thermalization Hypothesis \refs{\Deutsch\Srednicki-\Rigol,\LashkariVGJ} either in $2d$ \HikidaKHG\ or in higher $d$ using the approach of \MukhametzhanovZJA.

We also analyzed our bounds in the large $c$ theories with gravity duals \refs{\HartmanOAA, \WittenKT,\MaloneyUD,\BenjaminKRE}  and found \IntroMicroLargeC\ that the HKS result \HartmanOAA\ for the microcanonical entropy can be rigorously extended  to include the logarithmic correction with a bounded error of $O(1)$. In this case the microcanonical entropy counts  black hole microstates in AdS. In contrast to the situation in flat space, see e.g. \refs{\SenVZ,\MandalCJ}, the logarithmic correction to the black hole entropy in this case is universal and given by \LargeCentropy, as explained by Sen \SenDW. Our results for the logarithmic correction also agree with the old results of Carlip \CarlipNV\ (after averaging!) and constitute a rigorous derivation thereof.

It is important to emphasize that the techniques used in this paper require positivity of the spectral density $\rho(\Delta)$. Nothing of what we derived here holds if $\rho(\Delta)$ is not positive-definite. Even if the asymptotic of the partition function is fixed, one might imagine that many different spectral densities lead to the same asymptotic due to possible cancellations for a non-positive density. It is therefore not clear how to make rigorous the results of \KrausNWO, which involve (not necessarily positive) three-point functions.

Another important feature of our analysis is that the bounds that we obtained are in principle applicable at finite $\Delta$. To derive them we used the so-called HKS bound \HartmanOAA\ which allows one to estimate the contribution of heavy operators to the partition function. We have not fully explored these bounds and it would be interesting to do so, e.g. numerically. Recently a bound analogous to HKS was derived in the context of the four-point functions \KrausPAX. Therefore, it should be possible to repeat the large $\Delta$ conformal bootstrap analysis of \MukhametzhanovZJA\ at finite $\Delta$. 

One obvious extension of our analysis is to allow for non-zero angular potential. Since the combined spectral density $\rho(\Delta,J)$ is positive we should be able to derive the corresponding asymptotic results. The corresponding tauberian problem, however becomes two-dimensional. It will be interesting to extend our results to this case.

Most naturally our work should be thought of as a part of the modular bootstrap program \refs{\HellermanBU,\FriedanCBA\CollierCLS\CardyQHL-\ChoFZO} that systematically studies modular invariance by applying the most general set of linear functionals, both numerically and analytically. The functionals that appear in our work are particularly handy in deriving high-energy bounds. They are optimal in the sense that they give optimal scaling of error terms with $\Delta$ in the limit $\Delta \to \infty$, but not necessarily with optimal coefficients. In particular, it might be possible to improve the bounds \IntroMicroCardy, \sboundpm\ on $s(\delta, \Delta)$. It would be very interesting to find functionals that optimize these bounds in the spirit of \refs{\MazacQEV\MazacMDX-\MazacYCV}. We leave these tasks for the future.

\newsec{Acknowledgments}
 \noindent We would like to thank  A.~Belin, V.~Chandrasekaran, S.~Collier, A.~Dymarsky, D.~Jafferis, T.~Hartman, A.~R.~Levine, D.~Mazac, L.~Rastelli, J.~Sonner for useful discussions. The work of BM was supported in part by NSFCAREER grant PHY-1352084.

\appendix{A}{Estimate of $G(\nu)$}
\noindent In this appendix we derive the estimate \Gbound
\eqn\GboundA{
|G_\pm(\nu) | \leq 2 e^{-\beta \nu} \min[1, (\Lambda \nu)^{-2}] \ .
}
First, let's consider $G_+$ 
\eqn\Gp{
G_+(\nu) = {1\over 2\pi i } \int_{\beta - i \Lambda}^{\beta + i \Lambda} {dz \over z } {\Lambda^2 + (z- \beta)^2 \over \Lambda^2 + \beta^2 } e^{-\nu z} .
}

\ifig\contourA{Contour deformation for $G_+$.}{\epsfxsize2.7in\epsfbox{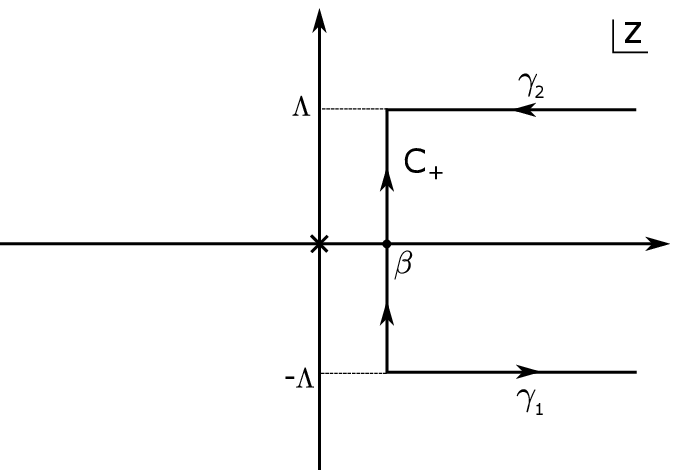}}

We will estimate \Gp\ in two different ways. First way is to deform the contour to the right to $\gamma_{1,2}$, see \contourA. Let's call the integrals over these contours $\Gamma_{1,2}$. Then we estimate as follows
\eqn\GammaestA{\eqalign{
|\Gamma_1| &= {1\over 2\pi} \left| \int_0^\infty {dx \over \beta - i \Lambda + x} { x(x-2i\Lambda) \over \Lambda^2 + \beta^2} e^{ - \nu(\beta - i \Lambda + x) } \right| \cr 
&\leq {e^{-\beta \nu} \over 2\pi (\Lambda^2 + \beta^2) \nu^2} \int_0^\infty dy~ e^{-y} y \sqrt{ {y^2/\nu^2} +4\Lambda^2 \over (\beta + y/\nu )^2 + \Lambda^2 } \ .
}}
Assuming $\nu>0$ the square root in the last integral is bounded by $2$. Therefore we have
\eqn\GammaestB{
|\Gamma_1| \leq {1\over \pi}{e^{-\beta \nu} \over  (\Lambda^2 + \beta^2) \nu^2}  \leq  {1\over 2} e^{-\beta \nu} (\Lambda \nu)^{-2} \ .
}
The contribution $\Gamma_2$ is bounded in the same way. In total we get
\eqn\GboundB{
|G_+(\nu)| \leq e^{-\beta \nu} (\Lambda \nu)^{-2} \ .
}
On the other hand we can estimate \GboundA\ as follows. First, we change variables in \GboundA\ $z = \beta + i \Lambda t$. After some algebra we get
\eqn\Gtvar{
G_+(\nu) =  {\Lambda^3 e^{-\beta \nu} \over \pi (\Lambda^2 + \beta^2)}   \int_0^1 dt ~ {(1-t^2)\over \beta^2 + \Lambda^2 t^2} \left[ \beta \cos (t \Lambda \nu)  - \Lambda t \sin (t \Lambda \nu) \right] \ .
}
Suppose $\Lambda |\nu|<1$. Then we estimate
\eqn\sinest{
|\beta \cos (t \Lambda \nu)  - \Lambda t \sin (t \Lambda \nu)| \leq \beta + \Lambda ^2 t^2 |\nu| \leq \beta + \Lambda t^2 \ ,
}
so that \Gtvar\ becomes
\eqn\GplusestA{\eqalign{
| G_+(\nu) | &\leq 
  {\Lambda^3 e^{-\beta \nu} \over \pi (\Lambda^2 + \beta^2)}   \int_0^1 dt ~ { \beta + \Lambda t^2  \over \beta^2 + \Lambda^2 t^2} \cr 
  & = {\Lambda^2 e^{-\beta \nu} \over \pi (\Lambda^2 + \beta^2)}  \left[ 1 + (1-\beta/\Lambda) \arctan\left( \Lambda /\beta\right) \right] \cr 
  & \leq  {\Lambda^2 e^{-\beta \nu} \over  \pi(\Lambda^2 + \beta^2)}  (1 + \pi /2) \leq e^{-\beta \nu} \ .
}}
Combining \GboundB, \GplusestA\ we get \GboundA\ for $G_+$.

Now we estimate $G_-$ for $\nu<0$. First, note that \GplusestA\ is valid for $\nu<0$ as well. Then we can use it to write
\eqn\GminA{
|G_-(\nu)| = |G_+(\nu) - 1| \leq  |G_+(\nu) | + 1 \leq 1 + e^{-\beta \nu} \leq 2 e^{-\beta \nu}, \qquad \nu<0 \ .
}

\ifig\contourB{Contour deformation for $G_-$.}{\epsfxsize2.7in\epsfbox{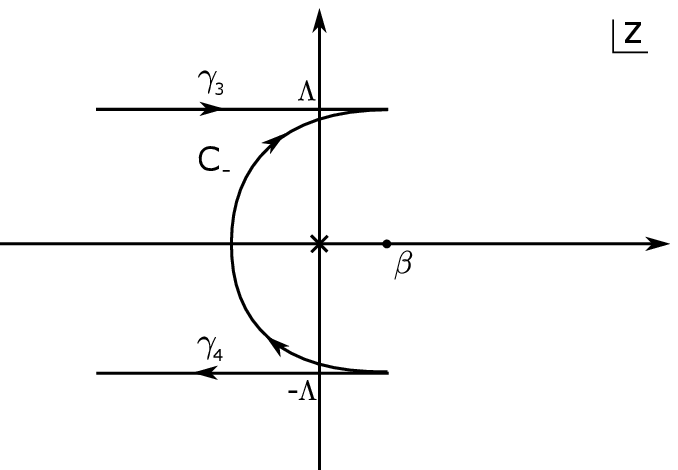}}

On the other hand we can deform the contour to the left and again get two contributions $\Gamma_{3,4}$, see \contourB. Then we estimate similarly to \GammaestA
\eqn\Gammamin{
|\Gamma_3| \leq {e^{-\beta \nu} \over 2\pi (\Lambda^2 + \beta^2) \nu^2} \int_0^\infty dy ~ e^{-y} y 
\sqrt{ {y^2/\nu^2} +4\Lambda^2 \over (\beta + y/\nu )^2 + \Lambda^2 } \ ,
} 
but now with $\nu<0$. Then we estimate for $\Lambda |\nu| >1$ 
\eqn\sqest{
\sqrt{ {y^2/\nu^2} +4\Lambda^2 \over (\beta + y/\nu )^2 + \Lambda^2 }  \leq 2 \sqrt{1 + {y^2 \over 4 \Lambda^2 \nu^2 }} \leq \sqrt{1 + {y^2 \over 4}}
}
so that \Gammamin\ becomes
\eqn\GammaminestA{
|\Gamma_3| \leq {e^{-\beta \nu} \over 2\pi (\Lambda^2 + \beta^2) \nu^2} \int_0^\infty dy ~ e^{-y} y \sqrt{1 + {y^2 \over 4}} \leq {3\over 2\pi}  {e^{-\beta \nu} \over (\Lambda^2 + \beta^2) \nu^2}  \leq e^{-\beta \nu} (\Lambda \nu)^{-2}
}
Making the same estimate for $\Gamma_4$ we finally get 
\eqn\Gminest{
|G_-(\nu)| \leq 2 e^{-\beta \nu} (\Lambda \nu)^{-2}
}
And combining this with \GminA\ gives \GboundA\ for $G_-(\nu)$.

\appendix{B}{Power Corrections}

\noindent We can consider multiple integrals of the density of states \MukhametzhanovZJA
\eqn\introMultiInt{\eqalign{
F_\rho^m(\Delta) &=  {1\over (m-1)! } \int_0^\Delta d\Delta' ~(\Delta - \Delta')^{m-1} \rho(\Delta')  = \cr 
 &= \int_0^\Delta d\Delta_{m} \int_0^{\Delta_{m}} d\Delta_{m-1} \dots \int_0^{\Delta_2} d\Delta_1 ~ \rho(\Delta_1) .
}} 
The tauberian theorem for \introMultiInt\ takes the form
\eqn\highermoments{\eqalign{
F_\rho^m(\Delta) &= {1\over (m-1)! } \int_0^\Delta d\Delta' ~(\Delta - \Delta')^{m-1} \rho_0(\Delta') + O\left( \Delta^{-3/4} e^{2\pi \sqrt{{c\over 3} \Delta} } \right) = \cr 
&=  {1\over 2\pi} \left( 3\over c\Delta \right)^{1/4} e^{2\pi \sqrt{{c\over 3} \Delta}} \left[ \sum_{i=1}^m c_i \Delta^{{i-1 \over 2}} + O(\Delta^{-1/2}) \right], \qquad \Delta \to \infty \ .
} }
The coefficients $c_i$ can be computed explicitly using the crossing kernel \naiveCardy. Again, all the spectral density moments $F_\rho^m(\Delta)$ are controlled by the unit operator in the dual channel. The intuition behind \introMultiInt\ is that each integration enhances smooth power-like terms while keeping intact oscillating non-universal terms. 

The derivation of \highermoments\ is analogous to the one in section 5. We consider \linfunc\ with a higher order pole
\eqn\linfuncpower{
{1\over 2\pi i }\int_{\beta - i \Lambda}^{\beta + i \Lambda} {dz \over z^{m+1}}  ~\left[ \Lambda^2 + (z - \beta)^2 \right] ~ e^{z \Delta } ~ \delta {\cal L}(z) \ .
}
When deforming the contour to the left, the pole contribution will produce the desired kernel $(\Delta - \Delta')^m$ from the expansion of $e^{z(\Delta - \Delta')}$ near $z=0$.\foot{Plus lower orders of $\Delta - \Delta'$ due to the expansion of the polynomial in \linfuncpower.}. The rest of the argument is identical to $m=0$ case.

\appendix{C}{Local bound at large c}

\noindent Let's estimate the sum 
\eqn\sparsesum{
A= e^{\beta \Delta}\sum_{\Delta_L \leq \Delta_H} \int_0^\infty d\Delta' ~ \rho_{\Delta_L} (\Delta') e^{-\beta \Delta'} \phi_+(\Delta') \ ,
}
where the crossing kernel of operator $\Delta_L$ is 
\eqn\crossL{
\rho_{\Delta_L}(\Delta) = 2\pi \sqrt{{c\over 12} - \Delta_L \over \Delta - {c\over 12}} 
I_1\left( 4\pi \sqrt{\left( {c\over 12} - \Delta_L \right) \left( \Delta - {c\over 12} \right)} \right) \theta(\Delta - c/12) + \delta(\Delta - c/12) \ . 
}
It reproduces the contribution of the operator $\Delta_L$ in the dual channel 
\eqn\crossLdef{
\int_0^\infty d\Delta ~\rho_{\Delta_L}(\Delta) e^{-\beta \Delta} = e^{- {4\pi^2 \over \beta} (\Delta_L - c/12)} \ .
}
Computing each integral by a saddle approximation we get 
\eqn\crossLL{\eqalign{
A = O\left( c^{-1/2} e^{2\pi c \sqrt{\e \over 3}} \sum_{\Delta_L \leq \Delta_H} e^{-4\pi \sqrt{3\e} \Delta_L} \right)
&= O\left( c^{-1/2} e^{2\pi c \sqrt{\e \over 3}} \right) \cr 
& = O(\rho_0(\Delta) ) , \qquad \e > {1\over 12} \ ,
}}
where we used the sparseness condition \HKSsparse.

\listrefs
\bye